\documentclass[aps,pra,twocolumn,10pt,superscriptaddress,floatfix,noeprint,nofootinbib]{revtex4-2}
\usepackage[T1]{fontenc}
\usepackage[latin9]{inputenc}
\usepackage{float}
\usepackage{mathrsfs}
\usepackage{leftindex}
\usepackage[caption=false]{subfig}
\usepackage{amsmath}
\usepackage{amssymb}
\usepackage{graphicx}
\usepackage{hyperref}
\usepackage[export]{adjustbox}
\hypersetup{colorlinks = true,
	linkcolor =blue,
	citecolor=blue, 
	urlcolor=blue 
}
\usepackage[normalem]{ulem}
\usepackage{xcolor}

\newcommand{\soutcolor}[2][red]{\bgroup\markoverwith{\textcolor{#1}{\rule[0.5ex]{2pt}{0.4pt}}}\ULon{#2}\egroup}

\usepackage{overpic}
\usepackage{xcolor}

\newcommand{\avg}[1]{\langle #1 \rangle}
\newcommand{\ket}[1]{\ensuremath{\vert #1 \rangle}}
\newcommand{\bra}[1]{\ensuremath{\langle #1 \vert}}
\newcommand{\mr}[1]{\mathrm{#1}}
\newcommand{\aop}{\ensuremath{\hat{a}}}
\newcommand{\adop}{\ensuremath{\hat{a}^{\dagger}}}
\newcommand{\mrd}{\mathrm{d}}
\newcommand{\Hop}{\hat{H}}
\newcommand{\Pop}{\hat{P}}
\newcommand{\Qop}{\hat{Q}}
\newcommand{\Rop}{\hat{R}}
\newcommand{\Aop}{\hat{A}}
\newcommand{\Rdop}{\hat{R}^{\dagger}}

\newcommand{\etaop}{\hat{\eta}}

\newcommand{\Eop}{\hat{E}}
\newcommand{\zetaop}{\hat{\zeta}}
\newcommand{\tg}{\tilde{g}}
\makeatletter
\AtBeginDocument{
	\addtolength{\abovedisplayskip}{-3mm}
	\addtolength{\abovedisplayshortskip}{-3mm}
	\addtolength{\belowdisplayskip}{-2mm}
	\addtolength{\belowdisplayshortskip}{-2mm}
}
\usepackage{bm}
\makeatother
\newcommand{\ldotsTwo}{%
	\mathinner{{\ldotp}{\ldotp}}%
}

\usepackage[english]{babel}

\begin{document}
	\title{Collective Dissipation of Oscillator Dipoles Strongly Coupled to 1-D Electromagnetic Reservoirs}
	\author{Subhasish Guha}
	\email{guhasubhasish@iitgn.ac.in}
	\affiliation{Indian Institute of Technology Gandhinagar, Palaj, Gujarat 382355, India}
	\author{Ipsita Bar}
	\email{ipsitab@iitgn.ac.in }
	\affiliation{Indian Institute of Technology Gandhinagar, Palaj, Gujarat 382355, India}
	\author{Bijay Kumar Agarwalla}
	\email{bijay@iiserpune.ac.in}
	\affiliation{Department of Physics, Indian Institute of Science Education and Research, Pune 411008, India}
	\author{B. Prasanna Venkatesh}
	\email{prasanna.b@iitgn.ac.in}
	\affiliation{Indian Institute of Technology Gandhinagar, Palaj, Gujarat 382355, India}
	\begin{abstract}
		We study the collective dissipative dynamics of dipoles modeled as harmonic oscillators coupled to 1-D electromagnetic reservoirs. The bosonic nature of the dipole oscillators as well as the reservoir modes allows an exact numerical simulation of the dynamics for arbitrary coupling strengths. At weak coupling, apart from essentially recovering the dynamics expected from a Markovian Lindblad master equation, we also obtain non-Markovian effects for spatially separated two-level emitters. In the so called ultrastrong coupling regime, we find the dynamics and steady state depends on the choice of the reservoir which is chosen as either an ideal cavity with equispaced, unbounded dispersion or a cavity array with a bounded dispersion. Moreover, at even higher coupling strengths, we find a decoupling between the light and matter degrees of freedom attributable to the increased importance of the diamagnetic term in the Hamiltonian. In this regime, we find that the dependence of the dynamics on the separation between the dipoles is not important and the dynamics is dominated by the occupation of the polariton mode of lowest energy.
	\end{abstract}
	\maketitle
	\section{Introduction}
	The study of collections of quantum emitters interacting with a common electromagnetic (EM) reservoir has emerged as an important paradigm of quantum optics and open many-body quantum systems \cite{Agarwal1973MasterEqnReview,gross_superradiance_1982,roy_colloquium_2017,chang_colloquium_2018,reitz_cooperative_2022}. Their relevance spans fundamental aspects of quantum optics underpinning phenomena such as collective spontaneous emission encapsulated by super- and subradiance \cite{gross_superradiance_1982}, collective optical response of atomic arrays and ensembles \cite{PhysRevA.94.043844,PhysRevLett.116.103602,PhysRevLett.118.113601,PhysRevLett.121.123606,bettles_quantum_2020,andreoli_maximum_2021} as well as important applications such as superradiant lasers with very small linewidths leading to precise atomic clocks \cite{meiser_prospects_2009,chen_active_2009,meiser_steady-state_2010,ostermann_protected_2013}, collective cooling of atoms \cite{domokos_collective_2002}, and long-lived quantum memories \cite{facchinetti_storing_2016,asenjo-garcia_exponential_2017,manzoni_optimization_2018}. Rapid development of diverse experimental platforms such as, trapped ions \cite{devoe_observation_1996,PhysRevLett.114.023602,begley_optimized_2016}, atomic gases \cite{guerin_subradiance_2016,PhysRevLett.117.073002,roof_observation_2016,ferioli_storage_2021}, atoms in optical lattices \cite{bohnet_steady-state_2012,norcia_superradiance_2016,norcia_frequency_2018} or tweezer arrays \cite{yan_superradiant_2023} placed inside optical cavities, and superconducting qubits coupled to cavities and waveguides \cite{lalumiere_input-output_2013,doi:10.1126/science.1244324,mirhosseini_cavity_2019,zanner_coherent_2022,sheremet_waveguide_2023}, to probe
	collective effects have further driven the research efforts in the field. The main theoretical tool used to describe collective dynamics of quantum emitters has been the Markovian Gorini-Kossakowski-Sudarshan-Lindblad master equation (referred to henceforth as the Lindblad master equation or master equation) adapted to the correlated dissipation scenario \cite{Agarwal1973MasterEqnReview,FICEK2002369}. This has been very successful in describing several experimental situations with weak coupling routinely realized in quantum optics. Nonetheless, this is a genuinely many body problem with the Hilbert space scaling as $l^{N_\mrd}$ where $l$ is the number of energy levels of the emitter and $N_\mrd$ the number of emitters. Thus, exact numerical simulations of the dynamics even with sophisticated methods can only go up to small $N_\mrd$ \cite{henriet_critical_2019}. As a result there is significant interest in developing analytical \cite{andreoli_maximum_2021} and numerical methods \cite{fux2024oqupypythonpackageefficiently,PhysRevA.93.043843} to tackle this challenging regime. 
	\begin{figure}
		\begin{centering}
			\begin{overpic}{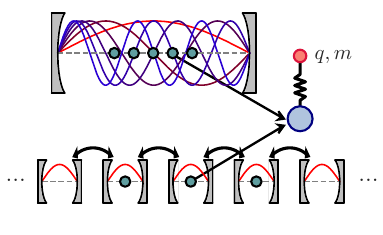}
				\put(0,30){\textbf{(a)}}
				\put(0,2){\textbf{(b)}}
			\end{overpic}
		\end{centering}
		\caption{Schematic diagram of an array of dipole oscillators interacting with the electromagnetic field inside (a) an ideal cavity and  (b) coupled cavity array.}
		\label{fig:schematic}
	\end{figure}
	
	Going beyond the weak coupling Markovian regime, the related problem of many emitters coupled principally to a single mode, in a sense the simplest structured reservoir, is the archetypal model of cavity quantum electrodynamics (cQED). Indeed strong coupling in such a setting has been studied extensively initially in the context of the famous phase transition in the Dicke Model \cite{PhysRev.93.99,PhysRevA.8.2517, PhysRevA.7.831,Garraway2011,kirton_introduction_2019}. Moreover, following the experimental realisation of the Dicke phase transition in \cite{baumann_dicke_2010}, attention returned to earlier works \cite{DickeNoGo1,DickeNogo2} which demonstrated that when the Hamiltonian describing the usual Dicke model is augmented by the terms proportional to the vector potential squared i.e. the so called $\mathbf{\Aop}^2$ terms, which cannot be neglected at strong coupling, lead to no-go theorems for the phase transition. Apart from helping understand the experiment \cite{baumann_dicke_2010} as a quantum simulation of the original Dicke model, this lead to a lively debate in the community clarifying various fundamental aspects of cQED such as the importance of gauge invariance \cite{Keeling_2007,Domokos2015, Domokos2012, Domokos2014, Nataf2010}, level truncation used to describe few-level quantum emitters \cite{Rabl_truncation2018}, as well as interesting phenomena in strong coupling cQED beyond the Dicke model \cite{Rabl_USC2016, Rabl_CQED2018, Rabl_CQED2023}. Furthermore, this and other experimental developments \cite{Niemczyk2010, Dietsche2011, Ebbesen2011, Gambino2014, Maissen2014, Zhang2014, Yoshihara2017, Bosman2017, Bayer_2017}, also heralded the studies of ultrastrong coupling regime and beyond in cQED \cite{Ciuti2005,Nori_USC2010,frisk_kockum_ultrastrong_2019,sheremet_waveguide_2023} where the coupling strength of the emitters to the EM field starts to approach the bare frequencies of the emitters and EM modes. In such extreme regimes of light-matter coupling several interesting features such as modification of the vacuum and associated phase transitions \cite{Rabl_USC2016, Rabl_CQED2018, Zueco2021,debernardis202}, excitations of multiple emitters via a single photon \cite{FrancoNori2016}, apparent superluminal signalling and its resolution from unavoidable multimode description \cite{Nori2018}, and eventually a decoupling of the light and matter degrees of freedom emerge \cite{PhysRevLett.112.016401,Ashida21,Ashida22}. 
	
	In this paper, our aim is to combine the two features highlighted so far and study the collective dissipation of emitters interacting strongly with a multimode EM environment. In this context there has already been an effort to study collective dissipation of quantum emitters beyond the Markovian regime \cite{Peter_Knight,ZhengBaranger2013b,Laakso2014,PhysRevA.96.043811,gonzalez-tudela_quantum_2017,sinha_collective_2020,sinha_non-Markovian_2020,Zueco2021,sheremet_waveguide_2023}. Nonetheless owing to both the strong coupling as well as the large Hilbert space, such studies are typically restricted for few quantum emitters, small number of excitations in the emitters, and additional restrictions such as the rotating wave approximation. As a tractable model without such restrictions to study collective dissipation with strong interactions, in this work we consider the exact dynamics of a collection of (quantum) harmonic oscillator dipoles interacting with one-dimensional (1-D) multimode EM reservoirs shown schematically in Fig.~\ref{fig:schematic}. While, the harmonic oscillators lack the non-linearity of few-level quantum emitters, the quadratic (in Bosonic operators) nature of the minimal coupling EM Hamiltonian allows for an exact numerical solution. We note that even though the dynamics of oscillators with bilinear coupling is classical, the initial conditions can be chosen as non-classical. Moreover, the study of QED with dipole oscillator emitters pioneered by Hopfield \cite{HopfieldModel} has also practical relevance for some set-ups such as intersubband excitations of semiconductor heterostructures coupled to microcavities \cite{Ciuti2005,anappara_signatures_2009}, Landau polaritons in two-dimensional electron gas coupled to cavities, and magnon-photon systems \cite{Hagenmueller2010,Zhang2014}. In addition, the exact solution also allows the possibility of tracking the dynamics of the EM reservoir modes during the collective dynamics providing a complementary view of the dynamics absent in the usual Lindblad master equation treatments.
	
	The rest of the paper and our results are organized as follows. In Sec.~\ref{sec:systemdescription}, we introduce the system Hamiltonian and the two specific 1-D EM reservoirs (cavity and cavity-array) of interest and also detail the numerical method for the exact calculation of the dynamics. Following this, we present the first set of results focusing on the regime with weak coupling between the dipoles and EM reservoir in Sec.~\ref{sec:WeakIntMarkov}. In this regime the standard Lindblad master equation description \cite{Agarwal1973MasterEqnReview} for collective dissipative dynamics is expected to hold allowing us to benchmark our numerical results. In addition we also uncover non-Markovian features in the dynamics that are absent in the Lindblad description \cite{sinha_non-Markovian_2020,PhysRevA.96.043811} for arrays of dipole oscillators. We move to the dynamics in the strong interaction regimes in Sec.~\ref{sec:stronginteraction} focusing first on the classification of the coupling regimes. This is followed by the presentation of the numerical results for the dynamics of both one and two dipole oscillators interacting strongly with the EM reservoir. A key result we find is that the strong coupling dynamics has features that depend on the nature of the EM reservoir (cavity vs cavity-array) both for one and two dipole oscillator cases. For the dynamics of two oscillators in the extreme coupling regimes (deep strong coupling and beyond) we find that the collective effects such as strong dependence of the dynamics on the separation between the dipoles are not important and the dynamics is dominated by the occupation of the polariton mode of lowest energy \cite{PhysRevLett.112.016401,Bayer_2017}. Furthermore, we find that similar considerations also extend to multiple oscillators in the strong coupling regimes. Finally, in Sec.~\ref{sec:conclusion} we summarize our central results and conclude the paper. Appendices \ref{app:A},\ref{app:B}, and \ref{app:C} provide some additional details omitted from the main text in Secs.~\ref{sec:systemdescription},\ref{sec:WeakIntMarkov}, and \ref{sec:stronginteraction}, respectively. 
	
	\section{System Description}
	\label{sec:systemdescription}
	In this section we will begin by presenting the details of our system of interest describing the interaction of dipole oscillators with a general one-dimensional electromagnetic (EM) reservoir. Following this we describe the specific EM reservoirs that we will consider in the paper, and finish with the key observables of interest of the system and the exact numerical approach we will use to calculate the same.
	
	\subsection{Hamiltonian}
	We consider a collection of oscillator dipoles interacting with a 1-D multimode electromagnetic environment (in a region of length $L$) with the Hamiltonian
	\begin{align}
		\hat{H}_{C}  &= \Hop_{\mr{dip}} + \Hop_{\mr{f}} + \Hop_{\mr{int}} + \Hop_{\mr{dia}}. \label{minimal_coupling}
	\end{align}
	In the above, the free Hamiltonian of the $N_\mrd$ oscillator dipoles (with mass $m$ and charge $q$) is given by (with $\hbar = 1$ henceforth)
	\begin{align}
		\Hop_{\mr{dip}} = \sum_{i=1}^{N_\mrd} \frac{\Pop_i^2}{2m} +\frac{m\omega_{si}^2 \hat{Q}_{i}^{2}}{2} \label{eq:FreeOscillatorHamiltonian} 
	\end{align}
	with $\omega_{si}$ denoting the frequency of the $i^{\mr{th}}$ oscillator and $\aop_i =  \frac{\sqrt{m \omega_{si}}}{2}\left[\Qop_i + i \Pop_i/(m \omega_{si})\right]$ ($\adop_i$) the corresponding annihilation (creation) operator. The free Hamiltonian of the electromagnetic environment that acts as the reservoir to which the dipoles are coupled is given by
	\begin{align}
		\hat{H}_{\mr{f}}=\sum_{n}\omega_{n}^{R}\hat{R}_{n}^{\dagger}\hat{R}_{n} \label{eq:FreeFieldHamiltonian},
	\end{align}
	with $\hat{R}_{n}$ ($\hat{R}_{n}^{\dagger}$) denoting the annihilation (creation) operators of the $n^{th}$ EM mode with frequency $\omega_n^{R}$.
	Note that in both Eqs.~\eqref{eq:FreeOscillatorHamiltonian} and \eqref{eq:FreeFieldHamiltonian} we have omitted the zero point contribution to the energy. The interaction Hamiltonian between the dipoles and the field and the diamagnetic term are given respectively by
	\begin{align}
		\Hop_{\mr{int}} &= \sum_{i=1}^{N_\mrd} -\frac{q}{m} \Pop_i \Aop (x_i), \label{eq:PdotAint}\\
		\Hop_{\mr{dia}} &= \sum_{i=1}^{N_\mrd} \frac{q^2}{2m} \Aop(x_i)^2,\label{eq:DiaMagTerm}
	\end{align}
	with $x_i$ representing the (classical) center of mass position of the $i^{\mr{th}}$ dipole oscillator and $\Aop(x_i)$ is the vector potential evaluated at the location of the oscillator (assuming the long wavelength limit). The vector potential can be expressed as:
	\begin{align}
		\mathbf{\Aop}(x) = \bm{\epsilon} \Aop(x)= \bm{\epsilon} \sum_{n}\mathscr{A}_{n}\left(\hat{R}_{n}f_{n}(x)+\hat{R}_{n}^{\dagger}f_{n}^{*}(x)\right)\label{Vector Potential},
	\end{align}
	where 
	\begin{align}
		\mathscr{A}_{n}=\sqrt{\frac{1}{2\epsilon_{0}\omega_{n}^{R}L \mathcal{A}}} \label{eq:Anfactor}
	\end{align}
	and $f_{n}(x)$ is the mode function corresponding to the mode of frequency $\omega_n^R$ \footnote{We assume that the mode functions are normalized as $\int_L \vert f_n(x) \vert^2 dx = L$}, and $\epsilon_0$ is the vacuum permittivity. Note that $\mathcal{A}$ denotes an area that in realistic quasi 1-D scenarios to be given by the 2-D cross-section area orthogonal to the propagation direction of the field and the (transverse) polarization of the EM field $\bm{\epsilon}$ is  taken to be along the direction of the oscillator displacements $\Qop_i$ for simplicity. Combining Eqs.~\eqref{eq:FreeOscillatorHamiltonian}, \eqref{eq:FreeFieldHamiltonian}, \eqref{eq:PdotAint}, and \eqref{eq:DiaMagTerm}, we can write the total Hamiltonian in the Coulomb gauge in terms of the annihilation and creation operators of the oscillator and EM field as
	\begin{align}
		\hat{H}_{C}  &= \sum_{i=1}^{N_\mrd} \omega_{si}\hat{a}_{i}^{\dagger}\hat{a}_{i}+\sum_{n}\omega_{n}^{R}\hat{R}_{n}^{\dagger}\hat{R}_{n}\nonumber\\
		& -i\sum_{i=1}^{N_\mrd} \sum_{n}\left(\hat{a}_{i}^{\dagger}-\hat{a}_{i}\right)\cdot\left(g_{in}\hat{R}_{n}+g_{in}^{*}\hat{R}_{n}^{\dagger}\right) \nonumber \\
		&+g_{0}^{2}\sum_{i=1}^{N_\mrd} \left[\sum_{n}\sqrt{\frac{1}{\omega_{n}^{R}}}\left(\hat{R}_{n}f_{n}(x_i)+\hat{R}_{n}^{\dagger}f_{n}^{*}(x_i)\right)\right]^{2},\label{H_C}
	\end{align}
	where 
	\begin{align}
		g_{0}&=q/\sqrt{4m\epsilon_{0}L \mathcal{A}}, \label{eq:g0defn}\\
		g_{in}&=g_{0}\sqrt{\omega_{si}/\omega_{n}^{R}}f_{n}(x_i) \label{eq:gindefn}
	\end{align}
	define the couplings. We note that using the Power-Zienau-Woolley transformation \cite{Cohen} the Coulomb gauge Hamiltonian Eq.~\eqref{H_C} can be written in its dipole gauge form as
	\begin{align}
		\hat{H}_{D}&= \sum_{i=1}^{N_\mr{d}}\omega_{si}\hat{a}_{i}^{\dagger}\hat{a}_{i}+\sum_{n}\omega_{n}^{R}\hat{R}_{n}^{\dagger}\hat{R}_{n}\nonumber\\
		&+i\sum_{i=1}^{N_\mr{d}}\sum_{n}\left(\hat{a}_{i}^{\dagger}+\hat{a}_{i}\right)\left(\tg_{in}^{*}\hat{R}_{n}^{\dagger}-\tg_{in}\hat{R}_{n}\right)\nonumber \\
		& + \sum_{i,j=1}^{N_\mr{d}} \Omega_{ij}^{*} \left( \adop_i \adop_j + \adop_i \aop_j \right) + \,\,\mr{h.c.},\label{H_D-1}
	\end{align}
	where the coupling is given by $\tg_{in} =g_{0}\sqrt{\omega_{n}^{R}/\omega_{si}}f_{n}(x_{i})$ and the self-interaction term for the oscillator dipoles is given by 
	\begin{align*}
		\Omega_{ij} = \sum_{n} \frac{\tg_{in}^*\tg_{jn}}{\omega_{n}^R} + \sum_{nm}\sum_{l=1}^{N_\mr{d}} \frac{\mr{Im}(\tg_{in}^{*}\tg_{ln})\mr{Im}(\tg_{im}^{*}\tg_{lm})}{\omega_{n}^{R}\omega_m^R \omega_{sl}}.
	\end{align*}
	Moreover, by ignoring the self interaction terms and performing a rotation of the reservoir operators $\Rop_{n}=-i\Rop_{n}$ in Eq.~\eqref{H_D-1} we obtain the quantum optical dipole interaction Hamiltonian as
	\begin{align}
		\hat{H}_{d} & = \sum_{i=1}^{N_{\mr{d}}}\omega_{si}\hat{a}_{i}^{\dagger}\hat{a}_{i}+\sum_{n}\omega_{n}^{R}\hat{R}_{n}^{\dagger}\hat{R}_{n} \label{eq:QOdipoleH} \\
		&+\sum_{n}\sum_{i=1}^{N_\mrd}\left(\hat{a}_{i}^{\dagger}+\hat{a}_{i}\right)\left(\tg_{in}\hat{R}_{n}+\tg_{in}^{*}\hat{R}_{n}^{\dagger}\right) \nonumber.
	\end{align}
	The above Hamiltonian is the starting point in derivations of the standard Lindblad master equation describing spontaneous emission of oscillators \cite{Agarwal1973MasterEqnReview}. In the above equations we have left the number of modes of the electromagnetic reservoir unspecified to allow for the general case of describing a continuum with infinitely many modes. But, throughout our numerical calculations below we will restrict to a finite number of modes, $N$, of the reservoir. Thus, the general problem we solve is the interacting dynamics of $N_{\mr{tot}} = N+N_\mrd$ oscillators. For the sake of completeness, we show that the normal mode frequencies of the system described by the dipole gauge Hamiltonian $\Hop_D$ in Eq.~\eqref{H_D-1} is exactly equal to the one described by the Coulomb gauge Hamiltonian $\Hop_C$ in Eq.~\eqref{H_C} (and that of the quantum optical Hamiltonian $\Hop_d$ in Eq.~\eqref{eq:QOdipoleH} at weak coupling) in Appendix \ref{app:A} \cite{frisk_kockum_ultrastrong_2019,Terradas2024}. Consequently, in the rest of the paper we will work with the Coulomb gauge Hamiltonian $\Hop_C$ in Eq.~\eqref{H_C}. 
	
	\subsection{Reservoir Description}
	
	While the above description of our set-up of interacting oscillators in principle can accommodate EM reservoirs in arbitrary spatial dimensions and with different spectra, for the sake of specificity in this work we restrict ourselves to two kinds of 1-dimensional EM reservoirs. The first one is a finite version of the 1-D continuum modelled as the field between two perfectly reflecting mirrors placed a distance $L$ apart i.e. a 1-D cavity. This is particularly suited to examine how the usual Lindblad master equation description of 1-D collective effects \cite{Agarwal1973MasterEqnReview,Peter_Knight,lalumiere_input-output_2013,Hill2017} emerges from the exact calculation considered here. Note that in this case the spectrum is equispaced and unbounded and we have to introduce an ultraviolet cut-off in our numerical calculations. The second kind of reservoir we will consider is a 1-D array of cavities that has a bounded spectrum and is a typical choice for studies of strong light-matter coupling \cite{Ashida21,Ashida22}.
	
	For a cavity in 1-D with mirrors at $x=-L/2$ and $x=L/2$, the spectrum is given by:
	\begin{align}
		\omega_n^R = n \omega_c = n \frac{c \pi}{L}, \,\, \,\  n\in \mathbb{Z}, \,\, 0<n \leq N\label{eq:SpectrumCavity}
	\end{align}
	with the free spectral range $\omega_c = c \pi/L$ providing the spacing, the frequency of the fundamental mode of the cavity, as well as the infra-red cutoff for the spectrum. We denote the ultraviolet cut-off as $\omega_{\mr{uv}} = N \omega_c$. The electric field inside the cavity are standing waves leading to the following form for the dimensionless mode function $f_n(x)$:
	\begin{align}
		f_n(x) = \begin{cases} 
			\sqrt{2} \sin\left( \frac{n\pi}{L}x\right) \,\, \mathrm{for}\,\, n \, \mathrm{even}\\
			\sqrt{2} \cos\left(\frac{n\pi}{L} x\right) \,\, \mathrm{for}\,\, n \, \mathrm{odd}
		\end{cases}
		\label{eq:ModeFnCavity}
	\end{align}
	For an 1-D array of $N$ cavities with resonance frequency $\omega_c$ and spacing $a$ the spectrum is given by \cite{Ashida21,Ashida22}
	\begin{align}
		\omega_n^R = \omega_c - J \cos \left( 2 \pi \frac{na}{L} \right), \,\, \,\, n \in \mathbb{Z}, -\frac{N}{2} \leq n < \frac{N}{2} \label{eq:SpectrumCavityArray}.
	\end{align}
	Here $L=Na$ and we take $N$ even for simplicity. The mode function for the cavity array takes the form
	\begin{equation}
		f_n(x) = e^{-i n \frac{2\pi}{L} x}
		\label{eq:ModeFnCavityArray}.
	\end{equation}
	
	\subsection{Observables and Exact Numerical Approach}
	
	Having described the Hamiltonian and the nature of the EM reservoirs, we are now in a position to describe the key observables we want to calculate to describe the collective dissipative dynamics of the dipole oscillators. The standard form of the initial condition of interest throughout the paper will have the oscillator dipoles in an excited state and the reservoir modes in their respective vacuum states.  Starting with this state we will track the dynamics of the dipole oscillator system by the average excitation number or population of the $i^{\mr{th}}$ dipole given as
	\begin{align}
		n_i(t) = \avg{\adop_i(t) \aop_i(t)} = \avg{\adop_i \aop_i} (t) \label{eq:populationdipoles},
	\end{align}
	as well as the total excitation in the system given by
	\begin{align}
		N_{\mr{exc}}(t) = \sum_{i=1}^{N_{\mr{d}}} \avg{\adop_i \aop_i} (t).\label{eq:totalexc}
	\end{align}
	In addition, in the weak coupling regime, the total radiated intensity can also be read off from the dynamics of the system excitation as 
	\begin{align}
		I(t) = -\omega_s \frac{dN_{\mr{exc}}(t)}{dt}.\label{eq:radintensity}
	\end{align}
	Moreover, given that we will be calculating the total dynamics of the dipole oscillators and the reservoir modes, the behavior of the electromagnetic modes can also be directly tracked using the intensity distribution of the electric field given by
	\begin{align}
		I\left(x,t\right)= \epsilon_{0} \avg{\Eop^{-}(x,t) \Eop^{+}(x,t)},
		\label{eq:IntensityofRadn}
	\end{align}
	where the positive and negatively rotating components of the electric field are defined by
	\begin{align}
		&\mathbf{E}\left(x,t\right) =  \bm{\epsilon} \left[ \Eop^{+}(x,t) + \Eop^{-}(x,t) \right] \nonumber \\
		&= i \bm{\epsilon} \sum_{n=1}^{N}\ \mathscr{E}_{n}\left[\Rop_{n}(t) f_{n} \left(x\right)-\Rdop_{n}(t)f_{n}^{*}\left(x\right)\right]     \label{eq:ReservoirEfield}
	\end{align}
	with $\mathscr{E}_{n}=\omega_{n}^{R}\mathscr{A}_{n}$, with $\mathscr{A}_{n}$ defined in Eq.~\eqref{eq:Anfactor}. In the case of the EM reservoir given by a cavity array, a better measure of the spatial distribution and dynamics is provided by the the photon number at each individual cavity site which is given by
	\begin{align}
		\avg{\hat{R}_x^{\dagger}(t)\hat{R}_x(t)} &= \avg{\hat{R}_x^{\dagger}\hat{R}_x} (t)\nonumber\\ &= \frac{1}{N}\sum_{kk^{\prime}} \avg{\hat{R}_{k^{\prime}}^{\dagger}(t)\hat{R}_{k}(t)} e^{-i(k-k^{\prime})x}, \label{eq:CavityArrayPhotNumPosn}
	\end{align}
	where we have used the definition $\hat{R}_{x}(t)=\frac{1}{\sqrt{N}}\sum_{k} \hat{R}_{k}(t)e^{-ikx}$ for the position space cavity array operators. Since all of the observables of interest described above are precisely given by equal time two point correlations of the dipole oscillator and EM reservoir operators, we next describe the numerical method we will employ to calculate the same.
	
	The total Hamiltonian of interest Eq.~\eqref{H_C} is a quadratic form in terms of the Bosonic operators of the dipole oscillators ($\adop_i,\aop_i$) and the reservoir modes ($\Rop_n,\Rdop_n$). As a result, our strategy to calculate the dynamics of two point correlations is to first diagonalise the Hamiltonian to identify the normal mode operators denoted by $(\zetaop_k,\zetaop_k^\dagger)$. Expressing the bare dipole and reservoir operators in terms of these normal modes and using their trivial time dependence will then give us the dynamics of correlations that we seek. 
	
	To perform the diagonalization, we organize the $N_{\mrd}$ dipole and $N$ reserovoir operators into the following $2N_{\mr{tot}}=2(N_\mrd +N )$ column vector of operators 
	\begin{align}
		\bm{\etaop} = \left [ \hat{a}_1, \ldotsTwo \hat{a}_{N_{\mrd}}, \hat{R}_1, \ldotsTwo \hat{R}_N,\adop_1,\ldotsTwo \adop_{N_\mrd},\Rdop_1, \ldotsTwo \Rdop_N \right]^T,\label{eta}
	\end{align}
	and denote the components of the operator vector $\bm{\etaop}$ by $\etaop_i$. Note that $\eta_i^\dagger = \eta_{i+N_{\mr{tot}}}$. This allows us to write the Hamiltonian in Eq.~\eqref{H_C} in the form
	\begin{align}
		\Hop_{C}=\frac{1}{2}\bm{\etaop}^{\dagger} H \bm{\etaop} -\frac{1}{2}\mr{Tr}[H_{d}] \label{eq:HCMatForm}.
	\end{align}
	Here $H$ is a $2N_{\mr{tot}}\times 2N_{\mr{tot}}$ hermitian matrix containing all the relevant system frequencies and couplings. The task then is to diagonalize the matrix $H$ with a transformation such that the resulting normal modes also preserve the structure of the column vector $\bm{\eta}$ with $\eta_i^\dagger = \eta_{i+N_{\mr{tot}}}$. This is indeed a kind of Bosonic Bogoliubov transformation \cite{Serafini} and we provide further details of the diagonalization procedure following \cite{key-6} in Appendix \ref{app:A}. In summary, we can diagonalize the matrix $H$ as
	\begin{align}
		H=T^{\dagger}\Lambda T,
	\end{align}
	with a complex transformation matrix $T$ and the diagonal matrix $\Lambda = (D \oplus D)$ with $D_{kk} = \lambda_k$ housing the normal mode frequencies $\lambda_k$. The normal mode operators can also be collected into a column vector 
	
	\begin{align}
		\bm{\zetaop} = T \bm{\etaop}
		\label{eq:PolaritonDefinition}.
	\end{align}
	The elements of the column vector $\bm{\zetaop}_i = \zetaop_i$ satisfy $\zetaop_{i+\mr{Ntot}} = \zetaop_{i}^\dagger$ and the Hamiltonian can be written as
	\begin{align}
		\Hop_{C}=\sum_{i=1}^{N_{\mr{tot}}} \lambda_{i}\left(\hat{\zeta}_{i}^{\dagger} \hat{\zeta}_{i}+\frac{1}{2}\right)-\frac{1}{2}{\rm Tr}\left[H_{d}\right] \label{eq:HopDiagonal}.
	\end{align}
	Since the normal modes $\zetaop_i$ are obtained as a linear combination of oscillator dipole (matter) and EM reservoir (light) modes, they will be referred to henceforth as polaritons. With the diagonal form of the Hamiltonian the time evolution of the polaritonic modes takes the simple form $\zetaop_i(t) = \zetaop_i(0) e^{-i\lambda_i t}$ in the Heisenberg picture. 
	
	For the calculation of equal time correlations of the bare system (dipole/reservoir) operators of the form $\avg{\etaop_i \etaop_j}(t)$, we begin by first calculating the initial correlations of the polariton operators $\avg{\zetaop_k \zetaop_l}(0)$ using their relation to the system operators given by Eq.~\eqref{eq:PolaritonDefinition} and the initial correlations $\avg{\etaop_i \etaop_j}(0)$. Following this we evolve the polariton correlations in time using $\zetaop_i(t) = \zetaop_i(0) e^{-i\lambda_i t}$ and calculate $\avg{\zetaop_k \zetaop_l}(t)$. Using the inverse of the Bogoliubov transformation $A= T^{-1}$, we can immediately write down the correlations of interest $\avg{\etaop_i \etaop_j}(t)$. We provide an expression for the same in Appendix \ref{app:A}. Thus, we can calculate the exact dynamics of a system of dipole oscillators coupled with arbitrary strength to an 1-D EM reservoir exploiting the quadratic bosonic nature of the total Hamiltonian. We next present the results of our calculations beginning with the weak coupling regime. 
	
	\section{Weak Coupling Regime - Markovian Dynamics}\label{sec:WeakIntMarkov}
	
	We begin the presentation of results by restricting the oscillator-field coupling to the weak regime. This is the standard regime of typical quantum optics treatments \cite{Agarwal1973MasterEqnReview,breuer_theory_2002,gross_superradiance_1982} where a Lindblad master equation can accurately model the dynamics. We will compare the results from the exact numerical calculation in the weak coupling limit to those from the master equation. This comparison will serve as a benchmark of the exact calculation and also allow us to highlight several aspects of how non-Markovian dynamics can arise even with weak interactions \cite{PhysRevA.96.043811,sinha_non-Markovian_2020} due to retardation effects. We begin with a quick recap of the master equations that are valid in the weak coupling regime for the two EM reservoirs of interest with particular focus on how to approach such a regime in our exact calculation.
	
	\subsection{Lindblad Master Equation}
	
	The standard approach for deriving a Lindblad master equation \cite{Agarwal1974,gross_superradiance_1982,Cohen,breuer_theory_2002} begins with the system of interest interacting with a reservoir. Following the Born, Markov, and secular approximations which are valid for a system interacting weakly with a large reservoir with a quasi-continuous and flat energy spectrum and evolving slowly with respect to the fast frequencies of the system and reservoir, the Lindblad master equation results. This is a first order trace and positivity preserving equation for the reduced density matrix of the system. Since the steps leading to this equation are routine \cite{Agarwal1974,gross_superradiance_1982,Cohen,breuer_theory_2002} we directly present the equation for the two reservoirs of interest, and after this comment on the parameter regimes in which the exact solution can approach the ideal predictions of the Lindblad master equation.
	
	The master equation for a collection of $N_{\mrd}$ oscillators with identical frequency $\omega_s$ placed at positions $\{x_i\}_{i=1}^{N_{\mrd}}$ in a 1-D cavity with spectrum and mode functions given by Eqs.~\eqref{eq:SpectrumCavity} and \eqref{eq:ModeFnCavity} is of the form \cite{Agarwal1974,gross_superradiance_1982}
	\begin{align}
		\frac{d}{dt}\rho_s & = - i \sum_{j=1}^{N_{\mrd}} \omega_s \left[ \hat{a}_j^{\dagger}\hat{a}_j , \rho_s \right] -i \sum_{i\neq j = 1}^{N_{\mrd}} \Omega_{ij} \left[ \hat{a}_i^{\dagger}\hat{a}_j , \rho_s \right] \nonumber\\
		&+\sum_{i,j=1}^{N_{\mrd}} \Gamma_{ij} \; \Big[ 2 \hat{a}_i \rho_s\hat{a}_j^{\dagger} - \{\hat{a}_i^{\dagger}\hat{a}_j,\rho_s\}\Big] \label{eq:collectiveME},
	\end{align}
	where the collective spontaneous emission rate $\Gamma_{ij}$ and collective Lamb shifts $\Omega_{ij}$ (reservoir induced dipole interactions) are given respectively as \cite{lalumiere_input-output_2013},
	\begin{align}
		\Gamma_{ij} &= g_0^2\frac{L}{2c}\cos{\Big(\frac{\omega_s}{c}\Delta x_{ij}}\Big) \label{eq:cavGamij}\\
		\Omega_{ij} &= g_0^2\frac{L}{2c}\sin{\Big(\frac{\omega_s}{c} \Delta x_{ij}  \Big)} \label{eq:cavDelij},
	\end{align}
	with $\Delta x_{ij} = \vert x_i-x_j \vert$ denoting the oscillator separation. In the same vein for $N_{\mrd}$ oscillators placed at positions (cavities) $\{x_i\}_{i=1}^{N_{\mrd}}$ in a coupled cavity array with $N$ resonators with spectrum and mode functions given by Eqs.~\eqref{eq:SpectrumCavityArray} and \eqref{eq:ModeFnCavityArray} respectively takes the same form as Eq.~\eqref{eq:collectiveME} with the only difference being the form of the collective spontaneous emission rate and Lamb shifts which are given as
	\begin{align}
		\Gamma_{ij} &= \frac{g_0^2 N \cos{(k_s \Delta x_{ij}})}{\sqrt{J^2 - \Delta_{cs}^2}} \label{eq:cavarrayGamij}\\
		\Omega_{ij} &= \frac{g_0^2 N \sin{(k_s \Delta x_{ij}})}{\sqrt{J^2 - \Delta_{cs}^2}} \label{eq:cavarrayDelij},
	\end{align}
	with $k_s a= \cos^{-1}{\left(\frac{\Delta_{cs}}{J}\right)}$ and $\Delta_{cs} \equiv \omega_c - \omega_s$ (system-cavity detuning). Note that the above master equations are obtained by taking the reservoirs to be in their ground state with temperature $T=0$. 
	
	\subsection{Markovian Limit with Finite Reservoir}
	
	We would next like to identify the parameter regime in which the predictions from the exact numerical approach detailed in Sec.~\ref{sec:systemdescription} C and the Markovian Lindblad master equation \eqref{eq:collectiveME} are expected to match each other. To this end we first note that the Lindblad master equation is derived under the assumption of a reservoir with dense spectrum and flat density of states around the system's energy level \cite{Agarwal1974,gross_superradiance_1982,Cohen,breuer_theory_2002}. This condition, especially the denseness of the spectrum, is realizable only in the infinitely large reservoir limit $\left(L\rightarrow\infty\right)$ in principle. However for finite reservoirs, the exact dynamics can qualitatively agree with the Markovian decay \cite{Peter_Knight} as long as the emitter cannot `see' the edge of the reservoir. This means, we require the timescale of the recurrence effect due to the finite size of the reservoir, denoted henceforth as $t_{\mr{fin}}$, is much larger than the oscillator's decay time-scale $t_{\gamma}$, such that it can completely decay before any finite size effects kick in. Let us identify the two time-scales for the reservoirs we are interested in. 
	
	For a 1-D cavity of length $L$, a simple estimate of the finite-size time-scale is given by the length of the reservoir divided by the velocity of propagation within the reservoir i.e. $t_\mr{fin,C} = L/c$. On the other hand, from Eq.~\eqref{eq:cavGamij} taking the limit of $i=j$, we can identify the Markovian single oscillator intensity decay rate ($2 \Gamma_{ii}$) as
	
	\begin{align}
		\gamma_\mr{C} = \frac{2\bar{g}_0^2}{c} \label{eq:cavdecayrate},
	\end{align}
	where the reservoir-size independent coupling parameter is given as $\bar{g}_0 = g_0 \sqrt{L} =q/\sqrt{2m\epsilon_0 \mathcal{A}}$ (recall $g_0$ is defined in Eq.~\eqref{eq:g0defn}). With this we can write $t_\gamma = 1/\gamma_\mr{C}$ and the condition for Markovianity as 
	\begin{align}
		\frac{2\bar{g}_0^2}{c^2}\gg \frac{1}{L} \label{eq:MarkovCondCavForm1}.
	\end{align}
	Thus, for given finite cavity size, the smallest coupling strength $\bar{g}_0$ such that the above equation is satisfied would lead to Markovian decay dynamics. We would like to state this condition in a more practical form for our consideration in this work where ultimately we would like to explore stronger coupling regimes between the oscillator and reservoirs. For the cavity reservoir, from Eq.~\eqref{eq:gindefn}, we can see that the coupling strength to the $n^{\mr{th}}$ mode $g_n \propto \sqrt{\omega_s/\omega_n}$ and $\omega_n = n \Delta \omega$ and the coupling strength is suppressed naturally for high frequency modes justifying the ultraviolet cut-off $\omega_{\mr{uv}}$. We provide details of the exact choice of this cut-off used in the calculations in Appendix \ref{app:B}. In order to maximise the number of reservoir modes quasi-resonantly interacting with the system, we always choose the system frequency to be in the middle of the reservoir spectrum (Eq.~\eqref{eq:SpectrumCavity}) i.e. $\omega_s = N \Delta \omega/2$. Rewriting the cavity length in terms of the system frequency, Eq.~\eqref{eq:MarkovCondCavForm1} reduces to:
	\begin{align}
		\{\phi_{\mr{C}} \equiv \frac{\bar{g}_0^2}{c \omega_s} \} \gg \frac{1}{N \pi} \label{eq:MarkovcondnCavity},
	\end{align}
	where we have defined the dimensionless parameter $\phi_{\mr{C}} \equiv \bar{g}_0^2/(c \omega_s)$ that will be used henceforth to classify the coupling strength regime of interest for the cavity reservoir. Thus, we find that after choosing a small enough $\phi_{\mr{C}}$ to ensure weak coupling satisfying the Born approximation, it is indeed important that we take enough modes in the reservoir such that Eq.~\eqref{eq:MarkovcondnCavity} is satisfied.
	
	Turning to the cavity array reservoir, we first note that the propagation velocity is given by the group velocity corresponding to the spectrum at the wave number corresponding to the system's frequency Eq.~\eqref{eq:SpectrumCavityArray}
	\begin{align}
		v_s = Ja\sin(k_n a)\vert_{k_n = k_s} = a \sqrt{J^2-\Delta_{cs}^2} \label{eq:velocityCA}.
	\end{align}
	In contrast to the cavity case, the propagation velocity depends on the system frequency $\omega_s$ (via the detuning $\Delta_{cs}$). Writing the finite size time-scale as $t_\mr{fin,CA} = L/v_s$, and reading off the spontaneous intensity decay rate from Eq.~\eqref{eq:cavarrayGamij} as
	\begin{align}
		\gamma_{\mr{CA}} = \frac{2 \bar{g}_0^2}{v_s} \label{eq:cavarraydecayrate},
	\end{align}
	we can write the time scales condition $t_\mr{fin} \gg \{t_\gamma = \gamma_{\mr{CA}}^{-1}\}$ as 
	\begin{align}
		\frac{2 \bar{g}_0^2}{v_s^2} \gg \frac{1}{L}
		\label{eq:MarkovcondCAform1},
	\end{align}
	which is a simple analog of Eq.~\eqref{eq:MarkovCondCavForm1} with $\bar{g}_0 = g_0 \sqrt{Na}$. As in the cavity case, we again want the system frequency to be in the middle of the cavity array spectrum, as the non-trivial band edge effects  can also lead to breakdown of Markovianity \cite{PhysRevA.96.043811,gonzalez-tudela_quantum_2017}. Hence, throughout the rest of the paper we take $\omega_c = \omega_s$, which leads to $k_s a = \pi/2$ resulting in $v_s = Ja$ and the Markovianity condition reduces to Eq.~\eqref{eq:MarkovcondCAform1}
	\begin{align}
		\{\phi_{\mr{CA}} \equiv \frac{2\bar{g}_0^2}{J^2 a}\} \gg \frac{1}{N} \label{eq:MarkovcondnCavityArray},
	\end{align}
	where we have defined a dimensionless measure of the coupling strength as $\phi_{\mr{CA}} \equiv 2\bar{g}_0^2/(J^2 a)$. Note that for our choice of the system frequency, $J$ provides a lower bound for the system frequency as we require $\{\omega_s=\omega_c\}>J$ to ensure positive frequencies for the reservoir modes. Thus, as before we need a large number of modes to ensure Markovian behavior for a given choice of (weak) coupling. In addition, as we explain in more detail in Appendix \ref{app:B}, we find that taking a value of $J/\omega_c$ not much smaller than or close to $1$ (we take $J/\omega_c = 0.5$ in our calculations) also enables Markovian behavior to emerge.
	
	Note that the conditions we have derived above to avoid finite size effects have been expressed as lower bounds for the coupling between the oscillators and reservoir. Even after choosing $N$ as large as possible, subject to ensuring that we can simulate the exact dynamics, we will still need to choose a finite value for the dimensionless coupling $\phi_{\mr{C,CA}}$. In other words as long as $N$ is finite, there will always be discreteness in the reservoir spectrum and this will lead to oscillatory dynamics from the exact solution over time scales much smaller than $\gamma_{\mr{C,CA}}^{-1}$ that will never be present in dynamics from the Lindblad master equation where a coarse graining \cite{Cohen} is assumed. As we will see in the following sections to see perfect agreement with the Markovian master equation for a finite sized reservoir, we will need to do a further coarse graining of the predictions from the exact numerical solutions over such small time scales. 
	
	\subsection{Single Oscillator Spontaneous Decay}
	\begin{figure}
		\begin{overpic}[width=0.8\columnwidth]{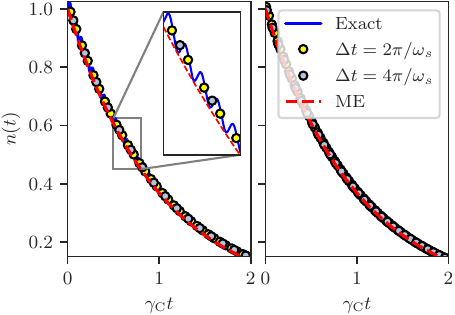}
			\put(20,20){\textbf{(a)}}
			\put(70,20){\textbf{(b)}}
		\end{overpic}
		\caption{Spontaneous decay of a single emitter in an ideal cavity of length $L=500 \pi c/\omega_s$, and $\omega_c=0.002\omega_s$ from the exact numerics (solid blue lines) and Lindblad master equation (dashed red lines) for $\phi_\mr{C}=0.01$ (a) and $\phi_{\mr{C}}=0.005$ (b). Inset in (a) highlights the difference between the exact solution without the coarse graining and the master equation. Colored dots are from coarse graining with different time-scales $\Delta t$. We have taken $N=1000$ modes of the cavity reservoir for the calculation.}
		\label{fig:1e_rad}
	\end{figure}
	\begin{figure}
		\begin{centering}
			\begin{overpic}[width=0.8\columnwidth]{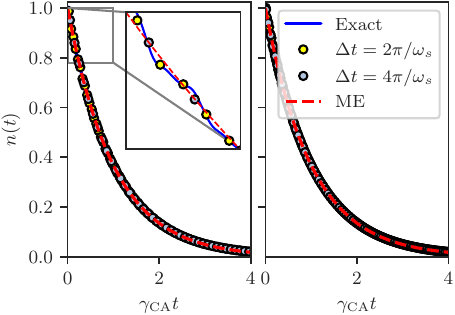}
				\put(20,20){\textbf{(a)}}
				\put(60,20){\textbf{(b)}}
			\end{overpic}
		\end{centering}
		\caption{Spontaneous Decay of a single emitter coupled to an array of $N=500$ coupled cavities with hopping parameter $J=1$ with $\omega_c=\omega_s=2J$. The coupling strength is chosen as $\phi_{\mr{CA}} = 0.04$ in (a) and $\phi_{\mr{CA}} = 0.02$ in  (b).Inset in (a) highlights the difference between the exact solution without the coarse graining and the master equation.  Colored dots are from coarse graining with different time-scales $\Delta t$.
			\label{fig:CCA_dec_1e}}
	\end{figure}
	In order to illustrate the approach to the Markovian limit with the exact simulations, we consider the simplest situation of a single oscillator, initialized in the Fock state $\ket{1}$, interacting with the reservoirs at vacuum. Focusing first on the cavity reservoir, in Fig.~\ref{fig:1e_rad} we present the results of the excitation number ($n(t) = \avg{\adop \aop} (t)$) dynamics of the oscillator for two values of the dimensionless coupling strength $\phi_{\mr{C}} = 0.01,0.005$ [(a) and (b) respectively] that are chosen to satisfy Eq.~\eqref{eq:MarkovcondnCavity} for the value of $N$ taken in the plots. We see that in both cases the results of the exact numerical calculation shows oscillatory dynamics about the Markovian predictions. The origin of this oscillatory dynamics, as alluded to before, arises ultimately from the discreteness of the reservoir as well as the stronger coupling to low frequency modes in the Coulomb gauge Hamiltonian. We discuss these oscillations further in Appendix \ref{app:B} and only note here that the frequency of the oscillation is set by the exchange dynamics of the system oscillator with low frequency modes of the reservoir and for our choice of system frequency in the middle of the reservoir spectrum it is of the same order as $\omega_s \gg \gamma_{\mr{C}}$. Thus, we need to perform an additional coarse graining, which practically involves taking a rolling or windowed average over a time scale $\Delta t \gtrsim \omega_s^{-1}$ on the results from the exact numerical solution to smoothen the oscillations. As evident from Fig.~\ref{fig:1e_rad}, such an averaging leads to very good agreement with the Markovian prediction as expected. Moreover we note by comparing Fig.~\ref{fig:1e_rad} (a) and (b), the oscillatory dynamics is less pronounced for smaller $\phi_\mr{C}$ and also requires smaller coarse graining timescale $\Delta t$. The results are similar even when the oscillators are initialized in coherent states (see Appendix \ref{app:B}). Note that we change the effective coupling $\phi_\mr{C}$ in Fig.~\ref{fig:1e_rad} by changing the bare coupling $\bar{g}_0$.
	
	In Fig.~\ref{fig:CCA_dec_1e}, we show that spontaneous decay of a single emitter coupled to the cavity array reservoir. As before we see that the agreement with the Markovian master equation can be controlled by the parameter $\phi_\mr{CA}$. In contrast to the cavity reservoir, we can see that the oscillations giving deviations from Markovianity are much less pronounced. This is in part due to the very different form of the density of states in the cavity and cavity array case. In the former, the density of states is independent of energy of the oscillator whereas in the case of the latter it takes the form $D(\omega) = N/(\pi \sqrt{J^2-(\omega_c-\omega)^2})$. Thus, the density of states at the system frequency $\omega_s = \omega_c$ can be controlled by an appropriate choice of the tunneling rate $J$ allowing Markovian behavior to be established (see Appendix \ref{app:B} for details). Having examined the emergence of Markovian behavior in the decay of single oscillators from the exact numerical simulation, in the rest of this section we will stick to the values of $\phi_\mr{C}$, $\phi_\mr{CA}$ and other system parameters from Figs.~\ref{fig:1e_rad} and \ref{fig:CCA_dec_1e} but consider collective dynamics with two or more oscillators. We will see that there are some unavoidable retardation induced non-Markovian effects that emerge in this case \cite{PhysRevA.96.043811,sinha_non-Markovian_2020,sinha_collective_2020}.
	
	\subsection{Two Oscillator Collective Dynamics}
	\begin{figure*}
		\begin{centering}
			\subfloat{\begin{overpic}[width=0.32\linewidth]{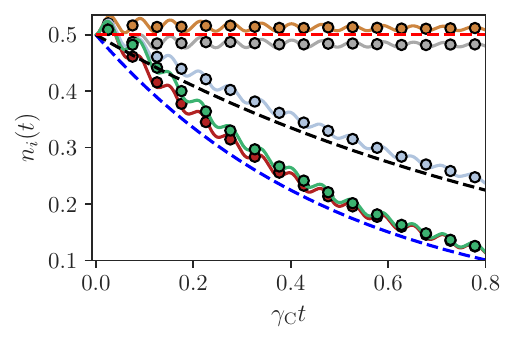}
					\put(65,50){\textbf{(a)}}
				\end{overpic}
			}
			\subfloat{\begin{overpic}[width=0.32\linewidth]{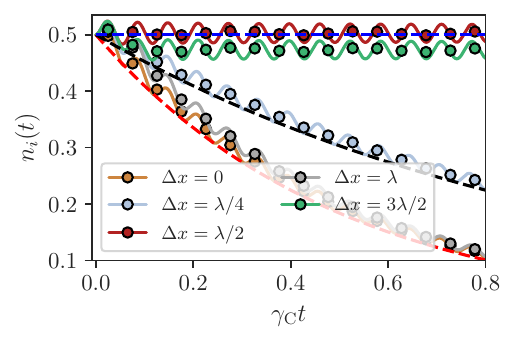}
					\put(65,50){\textbf{(b)}}
				\end{overpic}
			}
			\subfloat{\begin{overpic}[width=0.32\linewidth]{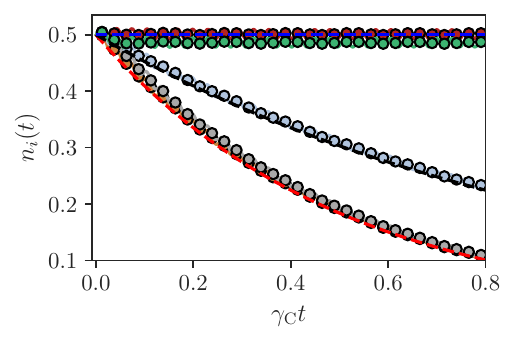}
					\put(65,50){\textbf{(c)}}
				\end{overpic}
			}
		\end{centering}
		\caption{Excitation number dynamics $n_i(t) = N_{\mr{exc}}(t)/2$ of two oscillator dipoles placed symmetrically about the center of and interacting weakly with a cavity
			plotted for different separations $\Delta x$ with the initial state $\ket{\psi_-}$ (a), $\ket{\psi_+}$ (b) with $\phi_\mr{C}=0.01$, and $\ket{\psi_+}$ with
			$\phi_\mr{C} = 0.005$ (c). Each plot displays the solution from the exact numerics (solid lines), coarse grained dynamics (circles) and the master equation (dashed lines). Other parameters are same as in Fig.~\ref{fig:1e_rad}.}
		\label{fig:2e_cav}
	\end{figure*}
	\begin{figure*}
		\begin{centering}
			\subfloat{\begin{overpic}[width=0.32\linewidth]{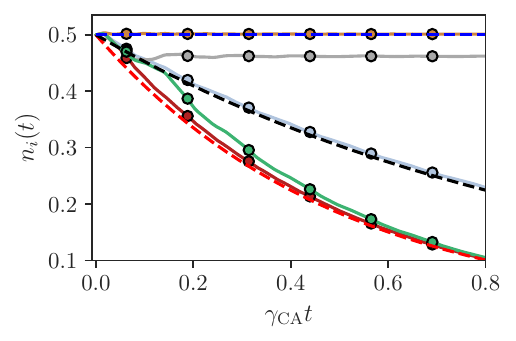}
					\put(65,45){\textbf{(a)}}
				\end{overpic}
			}
			\subfloat{\begin{overpic}[width=0.32\linewidth]{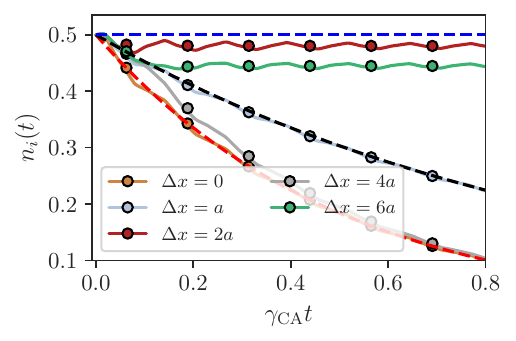}
					\put(65,45){\textbf{(b)}}
				\end{overpic}
			}
			\subfloat{\begin{overpic}[width=0.32\linewidth]{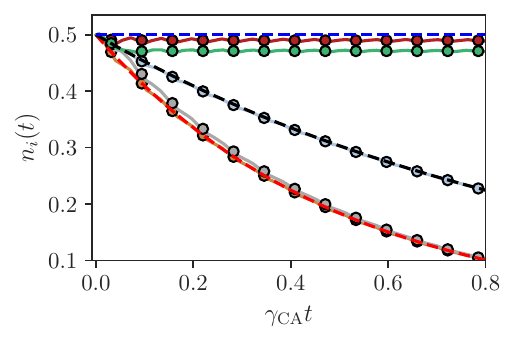}
					\put(65,45){\textbf{(c)}}
				\end{overpic}
			}
		\end{centering}
		\caption{Excitation number dynamics $n_i(t) = N_{\mr{exc}}(t)/2$ of two oscillator dipoles weakly interacting with a coupled cavity array with the initial state $\ket{\psi_-}$ (a), $\ket{\psi_+}$ (b) with $\phi_\mr{CA}=0.04$, and $\ket{\psi_+}$ with
			$\phi_\mr{CA} = 0.02$ (c). Predictions from the exact numerics (solid lines), coarsed grained dynamics (circular markers) and master equation (dashed lines) are depicted for different separations $\Delta x$. Other parameters are same as in Fig.~\ref{fig:CCA_dec_1e}.\label{fig:2e_CCA}}
	\end{figure*}
	As the first and simplest example of collective dissipative dynamics, let us now consider two oscillators interacting with the reservoirs. A natural starting point in discussions of superradiant dynamics of multiple oscillators is to take the initial state of the oscillators as a product state of the individual oscillators in either Fock or coherent states. The ensuing dynamics, for appropriate initial states \cite{Agarwal1974}, then leads to development of correlations and occupation of entangled states of the oscillators responsible for super or sub-radiant dynamics. We will indeed consider this situation in the next sub-section when discussing arrays of oscillators, but for the case of two oscillators there is a more expedient way to discuss collective dissipation. To that end let us consider the following correlated initial states of the oscillators:
	\begin{align}
		\ket{\psi_{\pm}} = \frac{1}{\sqrt{2}} \left( \ket{10} \pm \ket{01}  \right) \label{eq:supersub2osc},
	\end{align}
	with the notation $\ket{mn} = \ket{m} \otimes \ket{n}$. The above states are precisely the two oscillator super ($\ket{\psi_+}$) and sub-radiant ($\ket{\psi_-}$) states in the single excitation sub-space. Due to the secular approximation inherent in the Lindblad master equation, the dynamics of the oscillator dipoles with this initial state will be identical to that of two-level emitters and hence will allow us to make a direct comparison and interpret the results \cite{PhysRevA.96.043811,sinha_non-Markovian_2020}. 
	
	For the case of two oscillators and the initial condition Eq.~\eqref{eq:supersub2osc}, we can analytically solve the master equation \eqref{eq:collectiveME} and write down the total excitation number as
	\begin{align}
		N_{\mr{exc}}(t) = 2 n_i(t) =  e^{2(\mp \Gamma_{12}-\Gamma_0)t},
	\end{align}
	with $i=1,2$ and the total radiated intensity as
	\begin{align*}
		& I(t) =  2 \omega_s (\Gamma_0 \pm \Gamma_{12})e^{2(\mp \Gamma_{12}-\Gamma_0)t},
	\end{align*}
	with $\Gamma_0 = \Gamma_{11} = \Gamma_{22}$ and the $\pm$ denoting the two initial states $\ket{\psi_{\pm}}$. Note that $\Gamma_0 = \gamma_\mr{C}/2$ and $\Gamma_0 = \gamma_\mr{CA}/2$ for the cavity and cavity array reservoir respetively. As we can see, for $\Gamma_{12} = \Gamma_0$ ($\Gamma_{12}=-\Gamma_0$), the state $\ket{\psi_{+}}$ ($\ket{\psi_{-}}$) decays in a superradiant manner with a rate twice as fast as the independent oscillator decay rate $\Gamma_0$. In contrast for $\Gamma_{12} = -\Gamma_0$ ($\Gamma_{12} = \Gamma_0$) the state $\ket{\psi_{+}}$ ($\ket{\psi_{-}}$) is perfectly sub-radiant. From Eq.~\eqref{eq:cavGamij} we can immediately see that for the cavity reservoir the configuration
	\begin{align}
		\Delta x_{12} = r \lambda_s, r \in \mathbb{Z},
	\end{align}
	leads to superradiant (subradiant) emission for $\ket{\psi_+}$ ($\ket{\psi_-}$), and 
	\begin{align}
		\Delta x_{12} = (2r+1) \frac{\lambda_s}{2}, r \in \mathbb{Z},
	\end{align}
	leads to subradiant (superradiant) emission for $\ket{\psi_+}$ ($\ket{\psi_-}$). Here $\lambda_s = 2 \pi c/\omega_s$ denotes the wavelength of the cavity photon at the oscillator frequency. For 
	\begin{align}
		\Delta x_{12} =  (r+1/4) \lambda_s, r \in \mathbb{Z},\label{eq:indemitcondn}
	\end{align}
	we obtain $\Gamma_{12}=0$ and independent spontaneous emission. In the same manner, for oscillator frequency such that $\Delta_{cs} = 0$ and hence $k_s = \pi/(2a)$, in the cavity array scenario we can read off the superradiant (subradiant) configuration for $\ket{\psi_+}$ ($\ket{\psi_-}$) as 
	\begin{align}
		\Delta x_{12} = 4ra, r \in \mathbb{Z}
	\end{align}
	and subradiant (superradiant) configuration for $\ket{\psi_+}$ ($\ket{\psi_-}$) as
	\begin{align}
		\Delta x_{12} = 2ra, r \in \mathbb{Z}
	\end{align}
	from Eq.~\eqref{eq:cavarrayGamij}. For separations $\Delta x_{12}$ with 
	\begin{align}
		\Delta x_{12} = (2r+1)a, r \in \mathbb{Z},
	\end{align}
	we have $\Gamma_{12} =0$ leading to independent spontaneous emission. Note that the collective Lamb shifts $\Omega_{ij}$ do not make a difference to the excitation number and radiated intensity for the two oscillator scenario. 
	
	Having described the expected collective emission behavior for the two oscillator case from the Lindblad master equation, let us now compare the same to the exact numerical solution. In Fig.~\ref{fig:2e_cav} we show the oscillator excitation dynamics for two emitters starting initially in the states $\ket{\psi_\pm}$. While, overall, the behavior we have outlined above regarding the collective spontaneous decay is apparent, we see some important additional features in the exact dynamics. Most notably, there is a finite retardation time $t_\mr{ret}$ that is visible, most notably in Fig.~\ref{fig:2e_cav}(a), after which the exact dynamics prediction agrees with the master equation \cite{PhysRevA.96.043811,gonzalez-tudela_quantum_2017,sinha_collective_2020,sinha_non-Markovian_2020}. This retardation time, which is absent in the Markovian description, arises from the finite time required for the emitted radiation from the first emitter to reach the second emitter and can be estimated as the separation between the emitters divided by the speed of propagation of radiation
	inside the reservoir. As evident from Fig.~\ref{fig:2e_cav} (a,b), for $t<t_\mr{ret}$ the exact dynamics agrees with that of independent decay ($\Delta x_{12} = \lambda/4$) as over this timescale the emitters do not `see' each other. For the case of the cavity reservoir, the retardation time can be estimated as
	\begin{align}
		t_\mr{ret} \approx \frac{\Delta x_{12}}{c},
	\end{align}
	and for agreement of the exact simulation with the Markovian master equation results we require that the retardation time to be much less than the spontaneous decay timescale $t_\gamma = \gamma_C^{-1}$ i.e.
	\begin{align}
		\phi_\mr{C} \ll \left(2\pi \frac{\Delta x_{12}}{\lambda_s} \right)^{-1}.
	\end{align}
	Indeed as we can see from Fig.\ref{fig:2e_cav}(c) reducing $\phi_\mr{C}$ produces much better agreement with the Markovian master equation results. The results for two emitters coupled to different sites of a cavity array are presented in Fig.~\ref{fig:2e_CCA}. Much like the cavity reservoir case, there is overall agreement with the Markovian master equation for the weak values of $\phi_\mr{CA}$ we have chosen here except for the finite retardation time which can be estimated in this case as
	
	\begin{align}
		t_\mr{ret} \approx \frac{\Delta x_{12}}{Ja}.
	\end{align}
	The requirement that the above retardation time be much smaller than the decay timescale can be expressed as
	\begin{align}
		\phi_\mr{CA} \ll \left(\frac{\Delta x_{12}}{a} \right)^{-1}.
	\end{align}
	Similar to the cavity reservoir case, we see from Fig.\ref{fig:2e_CCA}(c) that reducing $\phi_\mr{CA}$ leads to much better agreement between the exact calculation and master equation dynamics. Thus, when considering the exact collective dynamics of two oscillators coupled to cavity and cavity array reservoirs an additional time scale $t_\mr{ret}$ becomes important. Indeed at a given weak value of the coupling strengths $\phi_\mr{C},\phi_\mr{CA}$ as the oscillators are separated by larger distances \cite{sinha_non-Markovian_2020,sinha_collective_2020} the non-Markovian effects due to this finite propagation time of excitations in the reservoir becomes significant.
	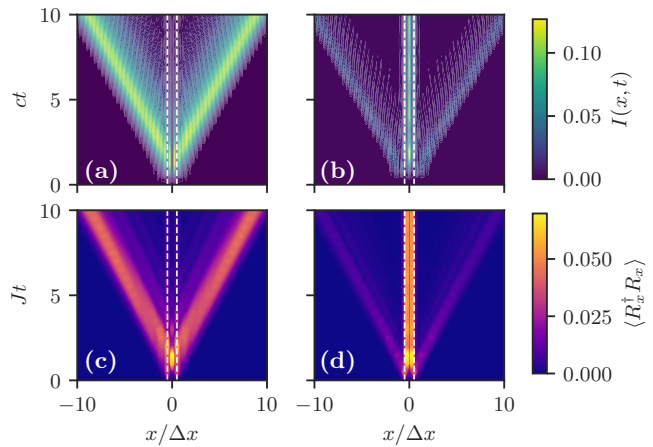
\begin{figure}
		\begin{centering}
			\begin{overpic}[abs,scale=.88]{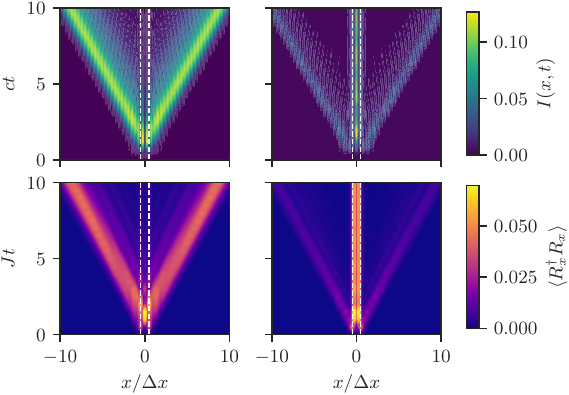}
				\put(28,102){\textbf{\textcolor{white}{(a)}}}
				\put(117,102){\textbf{\textcolor{white}{(b)}}}
				\put(28,30){\textbf{\textcolor{white}{(c)}}}
				\put(117,30){\textbf{\textcolor{white}{(d)}}}
			\end{overpic}
		\end{centering}
		\caption{Intensity distribution of electric field inside a cavity with $\phi_\mr{C} = 0.01$ (top panel) and photon number distribution in a coupled cavity array with $\phi_\mr{CA} = 0.04$ (bottom panel) for two emitters kept at a separation of $\Delta x = \lambda_s$ and $ \Delta x = 4a $  respectively, for an initially (a,c) superradiant and (b,d) subradiant state. Other parameters are same as in Figs.~\ref{fig:2e_cav} and \ref{fig:2e_CCA}. White dashed vertical lines denote the locations of the dipole oscillators.}
		\label{fig:reservoir_dyn}
	\end{figure}
	
	A nice way to visualize the collective superradiant and subradiant dynamics of two oscillators is by mapping out the dynamics of the reservoir modes which is possible from the exact simulation. To this end, for the cavity reservoir we calculate the electric field intensity $I(x,t)$ inside the cavity defined and the on-site photon number $\avg{\hat{R}_x^{\dagger}\hat{R}_x}(t)$ for the cavity array reservoir. Fig.~\ref{fig:reservoir_dyn} (a,b) shows the electric field intensity distributions of the cavity reservoir for two oscillators initialized in the state $\ket{\psi_+}$ with a separation of $\Delta x_{12} = \lambda$ leading to superradiant dynamics in (a) and $\Delta x_{12} = \lambda/2$ leading to subradiant dynamics in (b). Notice how in the subradiant dynamics the reservoir excitation is trapped in the region between the oscillators (positions are indicated by vertical white lines). In the subradiant scenario, we also see the effects of retardation as there is initial independent emission of light from the oscillators that spread out into the reservoir before the trapping behavior kicks in. In the same manner, Fig~\eqref{fig:reservoir_dyn} (c,d) shows the dynamics $\left\langle R_{x}^{\dagger}R_{x}\right\rangle$ as a function of $x$ and $Jt$ for two emitters coupled to two sites at a separation of $\Delta x_{12} = 4a$ of a cavity array. The initial states are chosen as $\ket{\psi_+}$ and $\ket{\psi_-}$ in (c) and (d) respectively leading to superradiant and subradiant dynamics.
	
	\subsection{Multiple Emitters}
	\begin{figure}
		\begin{centering}
			\subfloat{\begin{overpic}[width = 0.8\columnwidth]{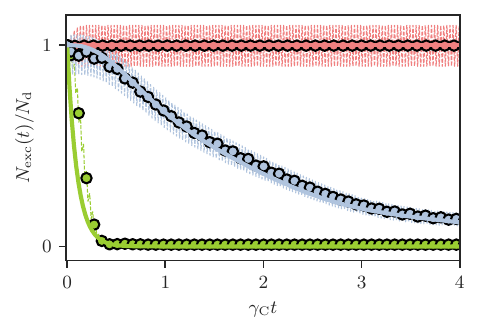}
					\put(80,30){\textbf{(a)}}
			\end{overpic}}\\
			\subfloat{\begin{overpic}[width=0.8\columnwidth]{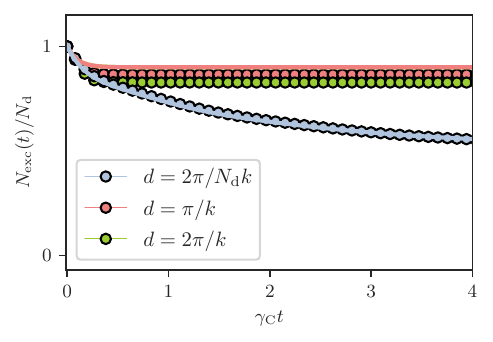}
					\put(80,30){\textbf{(b)}}
			\end{overpic}}
		\end{centering}
		\caption{Total excitation dynamics of $N_d=10$ oscillators placed symmetrically about the center inside a cavity with $\phi_{\mr{C}}=0.005$ in (a) coherent and (b) Fock initial states and  different separations $d$. Results from the coarse grained exact numerical calculations are (circles) compared with the Lindblad master equation results (solid lines). \label{fig:10e_pop}}
	\end{figure}
	
	We would now like to consider arrays of harmonic oscillators interacting with a cavity or cavity-array reservoir. In the context of two-level emitters, previous works have revealed a rich set of collective behavior that emerge from ordered arrays of quantum emitters coupled to a common electromagnetic reservoir \cite{lalumiere_input-output_2013,bettles_cooperative_2015,PhysRevA.94.043844,PhysRevLett.116.103602,asenjo-garcia_exponential_2017,masson_universality_2022,orell_collective_2022,holzinger_control_2022,cardenas-lopez_many-body_2023}. In the context of harmonic oscillator collective effects, while there have been general studies on collective dynamics \cite{katriel_algebraic_1970,agarwal_master-equation_1971,zakowicz_superradiant_1970,zakowicz_collective_1974,Agarwal1974,puri_master_1978,delanty_novel_2012} especially focusing on the comparison with the two-level emitters, the behavior of simple 1-D arrays have not been studied in much detail to the best of our knowledge. In what follows we examine the collective spontaneous emission dynamics of arrays of oscillators with mutual separation $d$.
	\begin{figure}
		\begin{centering}
			\subfloat{
				\begin{overpic}[width=0.8\columnwidth]{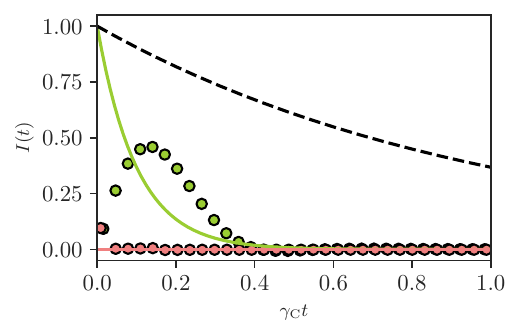}
					\put(80,45){\textbf{(a)}}
			\end{overpic}}\\
			\subfloat{\begin{overpic}[width=0.8\columnwidth]{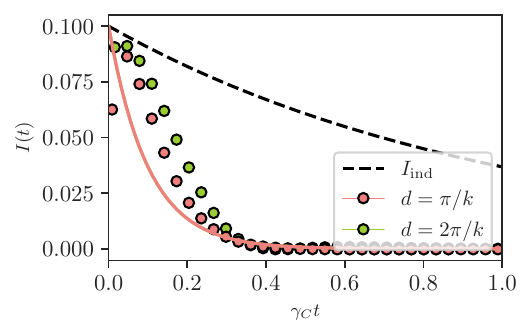}
					\put(80,45){\textbf{(b)}}
			\end{overpic}}
		\end{centering}
		\caption{Radiation rate of $N_d=10$ oscillators placed symmetrically about the center inside a cavity with $\phi_{\mr{C}}=0.005$ in (a) coherent and (b) Fock initial states and different separations $d$. Results from the coarse grained exact numerical calculations (circles) are compared with the Lindblad master equation results (solid lines) as well as independent oscillator radiation rate (dashed lines) scaled to allow comparison with the collective dynamics.  \label{fig:10e_rad}}
	\end{figure}
	We begin our analysis by recalling some exact, formal results regarding the spontaneous emission dynamics of multiple oscillators described by the collective master equation Eq.~\eqref{eq:collectiveME} originally presented in \cite{agarwal_master-equation_1971,Agarwal1974}. Writing the collective master equation using Glauber-Sudarshan P-function  leads to a linear Langevin equation allowing an exact solution for the dynamics. From this we find that the total oscillator excitation as a function of time can be written as 
	
	\begin{align}
		N_{\mr{exc}}(t) = \sum_{i,j} (e^{-G^\dagger t} e^{-G t})_{ij} \avg{\adop_i(0) \aop_j(0)}, \label{eq:ExcNumberArray}
	\end{align}
	with the matrix $G$ made of elements
	\begin{align}
		G_{ij} = \Gamma_{ij} + i (1-\delta_{ij})  \Omega_{ij},  \label{eq:Gmatrix}
	\end{align}
	where $\delta_{ij}$ is the Kronecker delta symbol. Furthermore, the intensity of radiation can be written as 
	\begin{align}
		I(t) = 2 \omega_s \sum_{ij} \left( e^{-G^\dagger t} \Gamma e^{-G t} \right)_{ij} \avg{\adop_i(0) \aop_j(0)} \label{eq:intensityarray},
	\end{align}
	with $\Gamma$ the matrix with elements $\Gamma_{ij}$. Further insight into the radiation intensity can be obtained by splitting it into a coherent and incoherent part \cite{agarwal_master-equation_1971} i.e. $I(t) = I_{\mr{incoh}}(t) + I_{\mr{coh}}(t)$ with
	\begin{align}
		I_{\mr{incoh}}(t) &= 2 \omega_s \sum_{i} \left( e^{-G^\dagger t} \Gamma e^{-G t} \right)_{ii} \avg{\adop_i(0) \aop_i(0)}\label{eq:Incohintensityarray} \\
		I_{\mr{coh}}(t) &= 2 \omega_s \sum_{i \neq j} \left( e^{-G^\dagger t} \Gamma e^{-G t} \right) \avg{\adop_i(0) \aop_j(0)} \label{eq:cohintensityarray}.
	\end{align}
	As we can see from the above definition, the coherent part of the radiation rate is non-zero only when the initial state of the oscillators is such that $\avg{\adop_i(0) \aop_j(0)} \neq 0$ for $i \neq j$. This can arise from taking special correlated initial states such as the many-emitter version of super and sub-radiant states $\ket{\psi_\pm}$ introduced earlier or for uncorrelated initial states with finite dipole moment i.e. $\avg{\adop_i(0) \aop_j(0)} = \avg{\adop_i(0)} \avg{\aop_j(0)} \neq 0$. A particular uncorrelated initial state with finite dipole moment is the one of the form $\ket{\psi(0)} = \bigotimes_{i = 1}^{N_\mrd} \ket{\alpha_i} =  \ket{\alpha_1(0)} \otimes \cdots \ket{\alpha_{N_\mrd}(0)}$ with each dipole in a coherent state $\ket{\alpha_i}$ (we will refer to such states henceforth as coherent initial states). In contrast for initial states with each oscillator in a Fock state ($\ket{n_i}$) of the form $\ket{\psi(0)} =  \bigotimes_{i=1}^{N_\mrd} \ket{n_i}$ (referred to henceforth as Fock initial states), the coherent term Eq.~\eqref{eq:cohintensityarray} is zero. Nonetheless, as we discuss below we can still get interesting collective dynamics and subradiant states for such initial states. In what follows, for simplicity, we will only consider Fock and coherent initial states and discuss the dynamics and steady-state predicted by the collective master equation for different ordered 1-D array configurations of oscillators interacting with cavity and cavity-array EM fields. We will also compare and contrast the results with the exact numerical solution. For the array configurations, we will consider oscillators arranged in an ordered array with spatial separation $d$ between consecutive oscillators.
	\subsubsection*{Coherent Initial State}
	For the coherent initial state, we can in-fact write the exact solution of the master equation \eqref{eq:collectiveME} as \cite{Agarwal1974}, $\hat{\rho}(t) = \ket{\psi(t)} \bra{\psi(t)}$ with $\ket{\psi(t)} =  \bigotimes_{i=1}^{N_\mrd} \ket{\alpha_i(t)}$ and
	\begin{align}
		\bm{\alpha} (t) = e^{-G t} \bm{\alpha}(0) \label{eq:coherentstatedynamicsarray},
	\end{align}
	and $\alpha_i(t) = \bm{\alpha}_i(t)$. Thus the collective master equation evolution Eq.~\eqref{eq:collectiveME} maps coherent product initial states to coherent states at later times. While we cannot analytically calculate the exponential in Eq.~\eqref{eq:coherentstatedynamicsarray} for 1-D ordered arrays of oscillators (interacting with a cavity or cavity-array fields) with arbitrary spacing $d$, we first show that we can make some strong statements regarding the steady-state. To this end, let us first note that as long as the real part of \textit{all} the eigenvalues of the symmetric matrix $G$ are non-negative and at least some are non-zero, we have that
	\begin{align}
		\bm{\alpha}(t\rightarrow \infty) = 0 \label{eq:sscoherent0}.
	\end{align}
	The semi-positivity of the eigenvalues of $G$ are guaranteed by the fact that the exact eigenvalues of the $\Gamma$ matrix for the 1-D array configuration are all zero except for two positive ones which are given by \cite{wolkowicz_bounds_1980}:
	\begin{align}
		\Gamma_{\pm} = \frac{N_\mrd \Gamma_0}{2} \pm \frac{\Gamma_0}{2} \frac{\sin(N_\mrd k d)}{\sin(kd)}, \label{eq:EmitRateArray1D}
	\end{align}
	with $k=\omega_s/c$ and $k=k_s$ for the cavity and cavity array cases respectively. Interestingly, we find that as long as $d \neq (2m+1) \pi/k$ or $d \neq m 2\pi/k$, the real part of \textit{all} the eigenvalues of the symmetric matrix $G$ are positive leading to the steady state Eq.~\eqref{eq:sscoherent0}. In particular, the steady state depends on the initial state for the special separations $d = (2m+1) \pi/k$ or $d = 2m\pi/k$ (with $m>0$ integer). We next discuss the dynamics and steady state for these special array separations as well as an additional array configuration given by $d = 2 \pi/(N_\mrd k)$.
	\begin{figure}
		\includegraphics[width=0.8\columnwidth]{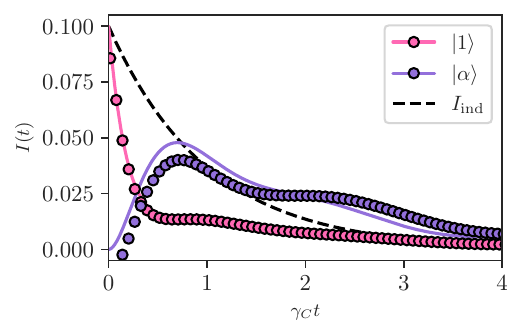}
		\caption{Radiation rates for an array of $N_d=10$ dipoles coupled with $\phi_{\mr{C}}=0.005$ to a cavity, with coherent (purple) and Fock (magenta) initial states in an array with separation $d=2\pi/N_\mrd k$. Coarse-grained exact dynamics (circles), Lindblad ME solution (solid lines), and independent oscillator radiation rate (dashed lines) are displayed for comparison. Other parameters are same as in Fig.~\ref{fig:10e_pop}
			\label{fig:10e_rad_lam10}}
	\end{figure}
	
	For the configurations $d = 2 \pi/k$  and $d = \pi/k$, the matrix $G$ has a simple form with $G_{ij} = \Gamma_0 \, \forall (i,j)$ and $G_{ij} = \Gamma_0 (-1)^{\vert i-j \vert}$ respectively and satisfies the relation $G^n = N_\mrd^{n-1}G$. This allows an exact evaluation of the exponential in Eq.~\eqref{eq:coherentstatedynamicsarray} to give
	\begin{align*}
		\bm{\alpha}(t) = \left[\bm{I} + \frac{\bm{J}}{N_\mrd}(e^{-N_\mrd \Gamma_0 t}-1) \right] \bm{\alpha}(0),
	\end{align*}
	with the matrix $\bm{J}$ having matrix elements $\bm{J}_{ij} = 1 \forall\{i,j\}$ for $d = 2 \pi/k$ and $\bm{J}_{ij} = (-1)^{\vert i-j \vert}$ for $d = \pi/k$. With this the total oscillator excitation and intensity become
	\begin{align*}
		N_\mr{exc}(t) &= \bm{\alpha}(0)^\dagger \left[\bm{I} + \frac{\bm{J}}{N_\mrd}(e^{-2N_\mrd \Gamma_0 t}-1) \right] \bm{\alpha}(0)\\
		I(t) &= 2 \omega_s \Gamma_0 \bm{\alpha}(0)^\dagger \bm{J} \bm{\alpha}(0) e^{-2 N_\mrd \Gamma_0t},
	\end{align*}
	respectively. For the simple scenario of each oscillator initialized in the same coherent state $\bm{\alpha}_i(0) = \alpha_0,\forall i$, we have 
	\begin{align*}
		N_\mr{exc}(t) &= N_\mrd \vert \alpha_0 \vert^2 e^{-2N_\mrd\Gamma_0t}\\
		I(t) &= 2\omega_s \Gamma_0 N_\mrd e^{-2 N_\mrd \Gamma_0 t},
	\end{align*}
	for the array with $d = 2 \pi/k$. Thus there is an enhanced (superradiant) decay of the initial excitation in this case. On the other hand for $d = \pi/k$, we get a perfect sub-radiant state with
	\begin{align*}
		N_\mr{exc}(t) &= N_\mrd \vert \alpha_0 \vert^2\\
		I(t) &= 0.
	\end{align*}
	These results are displayed in Figs.~\ref{fig:10e_pop} (a) and \ref{fig:10e_rad}(a) where we have also shown the comparison of the above Lindblad master equation predictions to the results of the exact simulations for oscillators placed in a cavity EM field. While there is overall agreement between the exact and Lindblad master equation, as in the two emitter scenario the exact numerics has fast oscillations and a coarse graining is needed to obtain good agreement (dots in Figs.~\ref{fig:10e_pop} and \ref{fig:10e_rad}). Since the decay rate scales as $N_\mrd \Gamma_0$ in situations with superradiant decay, we need to coarse grain on time scales satisfying $\{\Delta t \approx 2\pi/\omega_s\} \ll \left( N_\mrd \Gamma_0 \right)^{-1}$. Secondly, the exact solution always has a finite retardation timescale which is now governed by the distance between the dipoles which are placed the farthest apart. Thus $t_{ret}$ is proportional to $(N_\mrd -1)d$. This retardation effect is especially apparent in the plots for $I(t)$ in Fig.~\ref{fig:10e_rad}(a), where we see a large retardation time after which the coarse-grained dynamics give the fast-decay behavior similar to the Lindblad solution. Finally in Figs.~\ref{fig:10e_pop}(a) and Fig.~\ref{fig:10e_rad_lam10}, we have also shown the dynamics for $d = 2 \pi/(N_\mrd k)$. This value of $d$ is special as the two non-zero emission rates of the array given by Eq.~\eqref{eq:EmitRateArray1D} become equal i.e. $\Gamma_+ = \Gamma_- = N_\mrd \Gamma_0/2$ and the array shows global translational invariance \cite{Asenjo23}. Though at this point, the array of oscillators eventually decay to their ground state like the $d = 2 \pi/k$ configuration, the intensity of radiation shown in Fig.~\ref{fig:10e_rad_lam10}, is non-monotonic and slower than the decay rate of individual oscillators ($\sim \Gamma_0^{-1}$). 
	
	\subsubsection*{Fock Initial State}
	
	As before, the general solution, with Fock initial state for arrays with arbitrary separation, requires the exponentiation of the matrix $G$. Nonetheless, by arguments similar to the one we used for the coherent initial state, we can show that the ensemble of oscillators decays to a vacuum steady state for separations other than $d = 2m \pi/k$ and $d = 2 (m+1) \pi/k$. Focusing now on the specific case of $d = 2m \pi/k$ and $d = 2 (m+1) \pi/k$, we can use the fact the matrix $G$ simplifies to $G = \bm{J} \Gamma_0$ and hence can be exponentiated to calculate the key observables of interest $N_\mr{exc}(t)$ and $I(t)$ in Eqs.~\eqref{eq:ExcNumberArray} and \eqref{eq:intensityarray} respectively. To this end for the initial state $\ket{\psi(0)} =  \bigotimes_{i=1}^{N_\mrd} \ket{n_i}$ we obtain \cite{Agarwal1974,delanty_novel_2012}
	
	\begin{align*}
		N_\mr{exc}(t) = \bar{n}(0) \left[ N_\mrd - 1 + e^{-2N_\mrd\Gamma_0 t} \right],
	\end{align*}
	
	where $\bar{n}(0) = (\sum_i n_i)/N_\mrd$ for both the $d = 2m \pi/k$ and $d = 2 (m+1) \pi/k$ configurations. Consequently, the intensity of radiation becomes
	
	\begin{align*}
		I(t) = 2\omega_s \Gamma_0 N_\mrd \bar{n}(0) e^{-2N_\mrd\Gamma_0 t}.
	\end{align*}
	Comparing the above results to the coherent initial state for the $d=2 \pi/k$, we notice first that in both cases we do get an enhanced rate of decay $N_\mrd \Gamma_0$ but the steady state in the Fock initial state has residual excitation in the oscillators. In other words for the Fock initial state for both $d=2\pi/k$ and $d = \pi/k$ we have a sub-radiant steady state. These results are illustrated in Figs.~\ref{fig:10e_pop} (b) and ~\ref{fig:10e_rad} (b), and ~\ref{fig:10e_rad_lam10} where we see the dynamics predicted by the Lindblad evolution analysed above is as before compared with the exact numerical solution of the dynamics for oscillators placed in a cavity electromagnetic field. Apart from the caveats pointed out in the discussion regarding coherent initial state, which continue to hold even for the Fock initial state, we obtain good agreement with the exact solution. Finally, we find that qualitatively similar results also hold for oscillators placed in a cavity array and we present these results in Appendix \ref{app:B} for the sake of completeness.
	
	\section{Strong Interaction Regimes}
	\label{sec:stronginteraction}
	
	In this section we will consider the regimes of interaction, where the bare coupling between the system dipoles and the reservoir modes are no longer negligible compared to the the dipole's frequency. We have established in Sec.~\ref{sec:systemdescription} that the coupling between the $i^{th}$ dipole and the $n^{th}$ reservoir mode is given by $g_{in}$ [see Eq.~\eqref{H_C}]. In order to classify the coupling regimes and organize the discussion, we would like to define a single dimensionless parameter as done in \cite{frisk_kockum_ultrastrong_2019,PhysRevLett.112.016401} to label the strong coupling regimes. To this end, let us consider a dipole of frequency $\omega_s$ interacting with either a cavity or cavity-array reservoir. Omitting the index-$i$ in Eq.~\eqref{H_C}, the coupling of the dipole with the $n^{\mr{th}}$ reservoir mode can be written as $g_{n}=g_{0}\sqrt{\omega_s/ \omega_{n}^{R}}f_{n}\left(x\right)$. 
	
	For the cavity reservoir, frequency of the $n^{th}$ EM mode is $\omega^R_n = n \omega_c$, where $\omega_c=c\pi/L$ is the frequency of the fundamental cavity mode. Using this relation, we can write the coupling $g_n$ as \cite{PhysRevLett.112.016401}
	
	\begin{align*}
		g_{n}=\frac{g_{\mathrm{C}}}{\sqrt{n}}f_{n}\left(k_{n}x\right)
	\end{align*}
	where the $n$-independent cavity-dipole coupling strength $g_{\mathrm{C}}$ is given by
	\begin{align}
		g_{\mathrm{C}} = \bar{g}_0 \sqrt{\frac{\omega_s}{c \pi}},
	\end{align}
	where recall $\bar{g}_0 = g_0 \sqrt{L}$. Note that $g_{\mathrm{C}}$ is different from the bare coupling $g_0$ and can be interpreted as the coupling to the fundamental mode of an oscillator dipole placed at the center of a cavity. Since the coupling strength to the $n>1$ EM modes inside the cavity decreases by a factor of $\sqrt{n}$, $g_{\mr{C}}$ is also a measure of the largest possible cavity-dipole coupling in this multimode scenario. The coupling to the $n=n_{\mr{res}}$ resonant mode ($\omega_s = n_{\mr{res}} \omega_c$) is precisely given by the bare coupling $g_0=g_{\mathrm{C}}/ \sqrt{n_{\mr{res}}}$. As the Markovian treatment makes the broadband approximation, this coupling to the resonant mode determines the decay rate. However, we will see in the following subsections that the modes away from resonant energy will have the larger contributions to the population dynamics as we go beyond the Markovian regime. Hence, we define a dimensionless parameter, given by the ratio of the cavity-dipole coupling to the dipole oscillator frequency \textit{i.e.},
	\begin{align}
		\theta_{\mathrm{C}} = \frac{g_{\mathrm{C}}}{\omega_s} = \sqrt{\frac{\phi_{\mr{C}}}{\pi}},
	\end{align}
	which will henceforth be used to characterize the coupling regimes for the oscillator dipoles interacting with a cavity. Note that $\phi_{\mr{C}}$ was defined in Eq.~\eqref{eq:MarkovcondnCavity}.
	
	For a coupled cavity array, defining a dimensionless parameter starting from $g_n$ is more involved due to the non-linear form of the spectrum $\omega_n^R$. Following \cite{Ashida22}, we consider the expression $g_{\mr{CA}}^2 =\sum_{n=1}^N g_n^2$ to characterize the coupling between a single dipole oscillator and the many modes of the cavity array. In the limit of large $N$ we can approximate this (see Appendix \ref{app:C}) as
	\begin{align}
		g_{\mr{CA}} \simeq \sqrt{\frac{\bar{g}_0^2 \omega_s}{\pi J a}},
	\end{align}
	and write the dimensionless coupling parameter for the cavity array as
	\begin{align}
		\theta_{\mr{CA}} = \frac{g_{\mr{CA}}}{\omega_s} =& \sqrt{\frac{\phi_{\mr{CA}}J}{4\pi \omega_s}\cdot \frac{J}{\omega_s}},
	\end{align}
	with $\phi_{\mr{CA}}$ defined earlier in Eq.~\eqref{eq:MarkovcondnCavityArray}. Having identified the dimensionless parameters that characterize the coupling between the dipoles and the EM reservoirs, we next focus on the classification of the coupling regimes for different numerical values of $\theta_{\mr{C/CA}}$ based first on the behavior that can be discerned from the spectrum and the normal modes of the Hamiltonian. Following this we will highlight the different dynamical behavior we find in the strong coupling regimes for a single dipole as well as two dipole oscillators and conclude the section with a brief discussion analyzing and explaining our results.
	
	\subsection{Classification of Coupling Regimes}
	
	As described in Sec.~\ref{sec:systemdescription} C, since the Hamiltonian of our system of dipole oscillators (matter) interacting with an EM reservoir (light) is quadratic in terms of their respective Bosonic creation and annihilation operators, it can be diagonalized. The resulting normal mode frequencies $\{ \lambda_i \}$ and corresponding polaritonic operators $\{ \hat{\zeta}_i \}$ provide a clear way to characterize the coupled light-matter system. In this sub-section we exploit this to classify the different light-matter coupling regimes of the system as the dimensionless parameter $\theta_{\mr{C/CA}}$ is tuned. At infinitesimally \footnote{Here by infinitesimally small coupling, we mean coupling strengths smaller than the reservoir spectrum discretization. This is a regime which can be reached for any finite EM reservoir by lowering coupling strength as opposed to the case of continuous spectrum and infinite sized reservoir used in standard derivations of the Lindblad master equation.} small coupling, it is easy to see that the normal mode frequencies and the corresponding polaritonic operators will closely resemble the original undressed EM field and dipole oscillators. As $\theta_{\mr{C/CA}}$ is made small but finite, the polaritons display significant light-matter mixing or dressing initially around the frequencies resonant with the dipole oscillator leading to the (nearly) irreversible Markovian behavior discussed extensively in Sec.~\ref{sec:WeakIntMarkov}. As $\theta_{\mr{C/CA}}$ is made larger, the significant role of counter rotating terms in the atom-light interaction and eventually, as evident from Eq.~\eqref{H_C}, the increased importance of the diamagnetic term lead to two distinct regimes of strong light-matter interactions which we will classify, in what follows, using the associated behavior of the spectrum and polaritons. Such a classification of the different regimes of light-matter coupling via the static properties of the Hamiltonian will help us to anticipate and understand the dynamics in the different strong coupling regimes to be discussed in the next sub-section. 
	\begin{figure*}
		\subfloat{\begin{overpic}[width=0.32\linewidth]{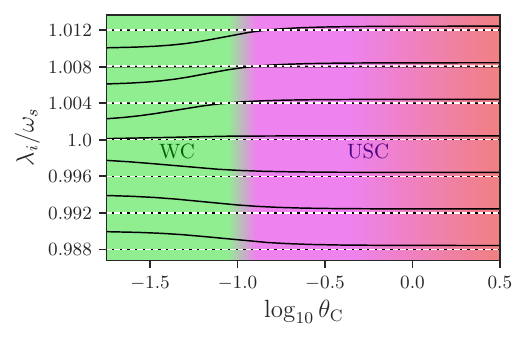}
				\put(70,55){\textbf{(a)}}
			\end{overpic}
		}
		\subfloat{\begin{overpic}[width=0.32\linewidth]{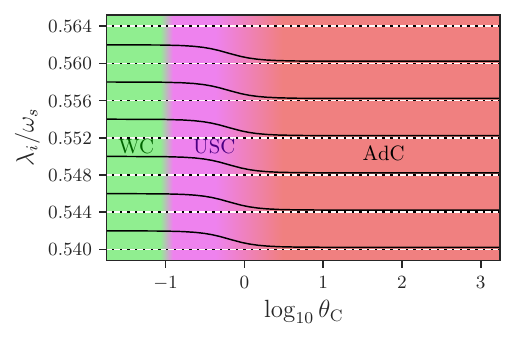}
				\put(70,55){\textbf{(b)}}
			\end{overpic}
		}
		\subfloat{\begin{overpic}[width=0.32\linewidth]{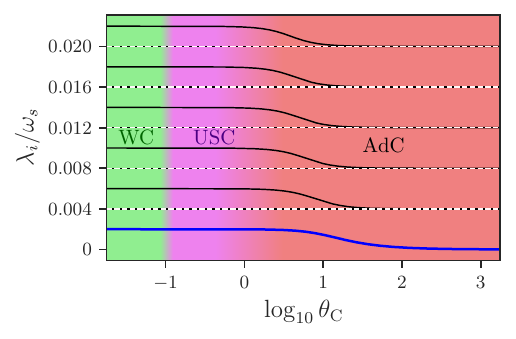}
				\put(70,55){\textbf{(c)}}
			\end{overpic}
		}
		\caption{Normal mode spectrum for a single dipole with frequency $\omega_s = 500 \omega_c$ in an ideal cavity as a function of the coupling parameter $\theta_\mr{C}$. Normal mode energies with $\lambda_i \sim \omega_s$ (a), $\lambda_i \sim \omega_s/2$ (b), and $\lambda_i \sim 0$ (c) are shown. Dashed lines represent uncoupled modes of the reservoir, and the solid blue line in (c) corresponds to the polariton mode with frequency tending to zero in the AdC regime.}
		\label{fig:classification}
	\end{figure*}
	\begin{figure*}
		\begin{centering}
			\subfloat{\begin{overpic}[width=0.32\linewidth]{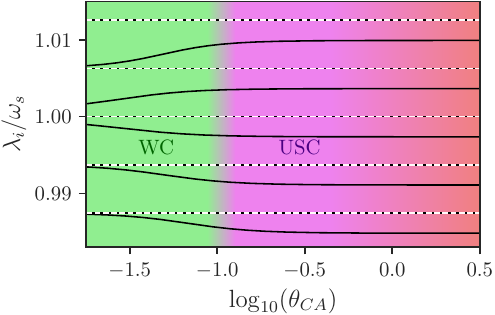}
					\put(70,58){\textbf{(a)}}
				\end{overpic}
			}
			\subfloat{\begin{overpic}[width=0.32\linewidth]{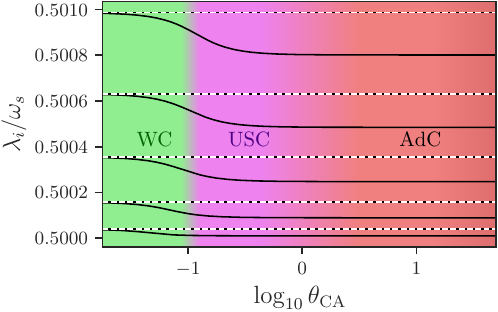}
					\put(70,58){\textbf{(b)}}
				\end{overpic}
			}
			\subfloat{\begin{overpic}[width=0.32\linewidth]{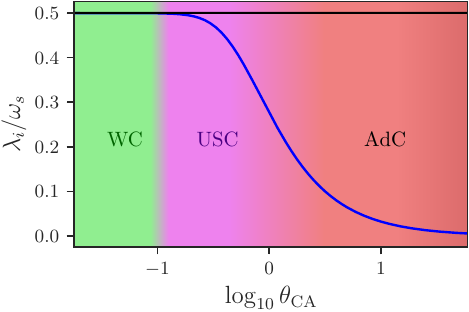}
					\put(70,58){\textbf{(c)}}
				\end{overpic}
			}
		\end{centering}
		\caption{Normal mode spectrum for a single dipole with frequency $\omega_s = \omega_c = 2 J$ coupled to a single site of a cavity array as a function of the coupling parameter $\theta_\mr{CA}$. Normal mode energies with $\lambda_i \sim \omega_s$ (a), $\lambda_i \sim \omega_s/2$ (b), and $\lambda_i \sim 0$ (c) are shown. Dashed lines represent uncoupled modes of the reservoir, and the solid blue line in (c) corresponds to the polariton mode with frequency tending to zero in the AdC regime.}
		\label{fig:spectrum_CCA}
	\end{figure*}
	
	\subsubsection{Spectrum} 
	
	The behavior of the normal-mode spectrum of the coupled dipole and EM reservoir can be broadly classified into three regimes as the parameter $\theta_{\mr{C/CA}}$ is varied. We will call these, in increasing order of $\theta_{\mr{C/CA}}$, as the weak coupling (WC), ultrastrong coupling (USC), and asymptotically decoupled (AdC) regimes. These regimes are depicted by green, purple, and red regions respectively in Fig.~\ref{fig:classification} (cavity reservoir) and Fig.~\ref{fig:spectrum_CCA} (cavity-array reservoir). 
	
	We denote the range of coupling $\theta_{\mr{C/CA}} \lesssim 10^{-1}$ as the weak coupling (WC) regime. At the edge of this regime with $\theta_{\mr{C/CA}} \ll 10^{-1}$, the spectrum resembles that of the uncoupled dipole and EM reservoir modes. In particular, for the cavity case we have considered a dipole placed at the center of the cavity $x = 0$ with frequency resonant with one of the cavity modes ($\omega_s = 500 \omega_c$ in Fig.~\ref{fig:classification} (a)). Here it is clear that for $\theta_{\mr{C/CA}} \ll 10^{-1}$, the spectrum is nearly equispaced with the same gaps of $\omega_c$ as the cavity mode spectrum with the single dipole oscillator mode almost degenerate with one of the cavity modes. Note that with this very weak coupling, for any finite sized reservoir with discrete spectrum, irreversible dynamics will not emerge. Moreover, due to the dipole's location at the center of the cavity, half of the modes (those with $f_n(x) \propto \sin(k_n x)$) are uncoupled to the dipole and are shown by dashed horizontal lines in Fig.~\ref{fig:classification} (see Appendix \ref{app:C} for details). As the parameter $\theta_{\mr{C}}$ is made larger but within the WC regime $\theta_{\mr{C/CA}} \lesssim 10^{-1}$, the hybridization between the EM field and dipole oscillator becomes significant. This is characterized by the lifting of the degeneracy between the dipole mode and the resonant cavity mode as well as significant modification of the energy levels \textit{near} the dipole's frequency as evident from Fig.~\ref{fig:classification} (a). Moreover, we note that the normal modes far away resonance still retain their photonic character as evident from their spectrum shown in Fig.~\ref{fig:classification} (b,c). As noted before, this is the regime where the irreversible Markovian dynamics (see Sec.~\ref{sec:WeakIntMarkov}) of the dipole oscillator emerges and in the language of Ashida et.al., \cite{Ashida21,Ashida22} the WC regime is essentially a perturbative regime of light-matter coupling.
	
	Coming to the cavity array case, the photonic reservoir's eigenmodes are travelling waves. In the absence of the dipole oscillator, these modes with frequency $\omega_{k_n}$ are all doubly degenerate (for any non-zero wave number $k_n$) since there are both left and right moving waves inside the array (see Eq.~\eqref{eq:SpectrumCavityArray}). Furthermore, coupling a dipole oscillator with frequency $\omega_s = \omega_c$ resonant with one of the eigenmodes leads to a (near) triple degeneracy around the resonant frequency for very weak coupling ($\theta_{\mr{CA}} \ll 1$) as shown in Fig.~\ref{fig:spectrum_CCA}(a). Moreover, just as in the cavity case, we can also show (see Appendix \ref{app:C}) that there are always a family of uncoupled, purely photonic modes that emerge (shown by dashed lines in Fig.~\ref{fig:spectrum_CCA}) even in the cavity array case. Apart from this at the edge of the WC regime with $\theta_{\mr{CA}} \ll 1$, the spectrum is essentially given by the photonic dispersion of the cavity array
	given in Eq.~\eqref{eq:SpectrumCavityArray} which is not equispaced unlike the cavity as evident from Fig.~\ref{fig:spectrum_CCA}(b). As in the cavity case, increasing $\theta_{\mr{CA}}$ within the WC regime leads to hybridization of light-matter degrees of freedom, lifting of degeneracy with the dipole mode and significant modification of the spectrum around the dipole oscillator's frequency that leads to emergence of Markovian dynamics. One significant difference to note in the cavity array case compared to the cavity is that deep in the WC regime ($\theta_{\mr{CA}} \lesssim 10^{-1}$) the frequency of modes away from the dipole oscillator frequency are also strongly modified as shown in Fig.~\ref{fig:spectrum_CCA} (a,b).
	
	As the coupling is further increased in the range $10^{-1} \leq \theta_{\mr{C/CA}} \leq 10^{0}$, we reach the Ultra-strong coupling (USC) regime \cite{frisk_kockum_ultrastrong_2019,Ashida22}. For the cavity case, this regime is characterized by strong modification of the normal mode frequencies of modes far away from the dipole oscillator frequency as depicted in the purple regions in Figs.~\ref{fig:classification} (b) where the normal mode frequencies around $\lambda \sim \omega_s/2$ have been shown. Similar results also hold for the cavity array as shown Fig.~\ref{fig:spectrum_CCA} (b) but with an important difference. In the USC regime for the cavity array, even the lowest energy modes (infrared region) around $\omega_{k_n} \sim \omega_c-J$ are modified whereas this is not the case for the cavity as can be seen in Fig.~\ref{fig:classification}(c). 
	
	Finally, we denote the the region with $\theta_{\mr{C/CA}} > 10^{0}$ as the asymptotically decoupled (AdC) regime  \cite{PhysRevLett.112.016401,Ashida22} and is shown in red in Fig.~\ref{fig:classification} (b,c). The justification for this naming will become apparent in the forthcoming sections where we will see that in this regime an effective decoupling of the light and matter degrees of freedom emerges. The key feature of the spectrum in this regime for the cavity case is the strong modification of even the lowest part of the spectrum (near the infrared cutoff of $\omega_c$) culminating with the normal mode of the smallest frequency asymptotically going to zero (thick blue line in Fig.~\ref{fig:classification} (c)) at very large $\theta_{\mr{C}}$. Finally, we note that at the largest values of $\theta_{\mr{C}}$ we have considered in Fig.~\ref{fig:classification} (c), the spectrum becomes equispaced again (with double the spacing of the cavity spectrum $\omega_c$) with the coupled polaritonic modes becoming degenerate with the uncoupled photonic modes. This suggests that in this regime, the cavity is split into two halves of equal length by the dipole placed at the center \cite{de_liberato_light-matter_2014}. We note that regimes with $\theta_{\mr{C}} \sim 1$ are known in literature as deep strong coupling \cite{PhysRevLett.105.263603,PhysRevLett.112.016401, Ashida21,Ashida22} and $\theta_{\mr{C}} > 1$ as extremely strong coupling \cite{Ashida21,Ashida22}. We do not specifically distinguish between these regimes and group them together as AdC regime since the dynamical properties of the system we consider do not change between these regimes. Turning our focus to the cavity array in the AdC regime with $\theta_{\mr{CA}}>1.0$, from Fig.~\ref{fig:spectrum_CCA} (c), we see that the lowest energy normal mode asymptotically goes to zero frequency again. Moreover, we can also see by comparing Figs.~\ref{fig:spectrum_CCA} (c) and \ref{fig:classification} (c) that the lowest energy normal mode approaches its asymptotic value faster in the cavity array case than in the cavity.
	
	In summary, we have classified the coupling regimes of the dipole oscillator interacting with a multimode EM field reservoir using the behavior of the normal mode spectrum. While the irreversible dynamical behavior in the WC regime was discussed extensively in Sec.~\ref{sec:WeakIntMarkov}, we will present results for the USC, AdC regimes in the following sections. Furthermore, while the behavior of the spectrum is qualitatively similar for the cavity and cavity-array reservoirs, we have also identified some interesting differences especially in the USC and AdC regimes which will again be reflected in the polariton behavior and dynamics. Our classification scheme for the USC and AdC is in agreement with previous works \cite{frisk_kockum_ultrastrong_2019,Ashida22} but we note that since our interest is in modelling EM reservoirs as explicitly multimode ones, we have not distinguished the standard strong coupling regime relevant to single-mode cavity QED. 
	\begin{figure}
		\begin{centering}
			\hspace{0.4cm}\subfloat{\begin{overpic}[width=0.8\columnwidth]{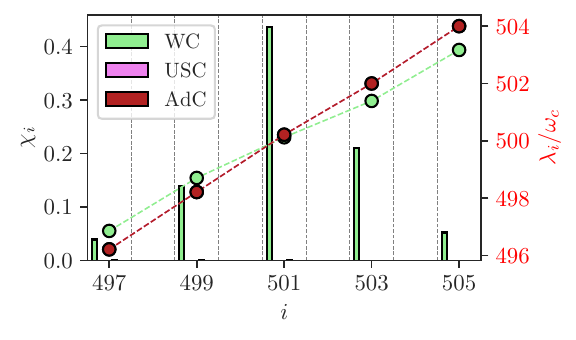}
					\put(70,30){\textbf{(a)}}
				\end{overpic}
			}\\
			\subfloat{\begin{overpic}[width=0.8\columnwidth]{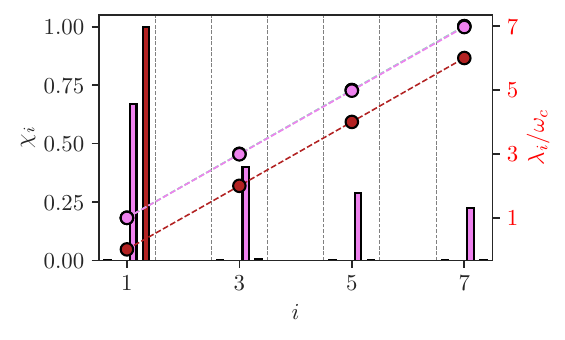}
					\put(70,30){\textbf{(b)}}
				\end{overpic}
			}
		\end{centering}
		\caption{Matter component of the $i^{\mr{th}}$ polariton mode (bar plots, left axis) for a single dipole oscillator placed at the center of an ideal cavity in the weak, ultrastrong, and asymptotically decoupled regimes with $\log_{10}\theta_{\mr{C}}=(-1.5,-0.1,2)$ respectively. Circular markers represent the normal mode frequency of the $i^{\mr{th}}$ polariton (right axis). Panel (a) and (b) respectively show the behavior of polariton modes with frequency resonant with the dipole oscillator of frequency $\omega_s = 500 \omega_c$ and with small frequency.}
		\label{fig:dressing}
	\end{figure}
	\begin{figure}
		\begin{centering}
			\hspace{0.4cm}\subfloat{\begin{overpic}[width=0.8\columnwidth]{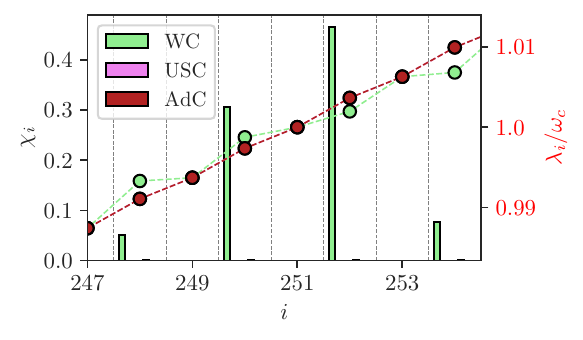}
					\put(70,30){\textbf{(a)}}
				\end{overpic}
			}\\
			\subfloat{\begin{overpic}[width=0.8\columnwidth]{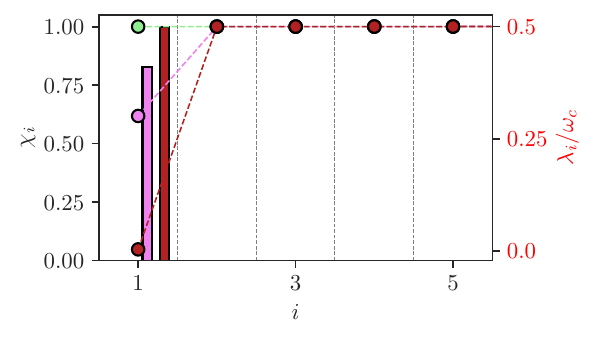}
					\put(70,30){\textbf{(b)}}
				\end{overpic}
			}
		\end{centering}
		\caption{Matter component of the $i^{\mr{th}}$ polariton mode (bar plots, left axis) for a single dipole oscillator of frequency $\omega_s = \omega_c$ coupled to a single site of a cavity array in the weak, ultrastrong, and asymptotically decoupled regimes with $\log_{10}\theta_{\mr{CA}}=(-1.65,-0.05,2)$ respectively. Other plot details remain the same as Fig.~\ref{fig:dressing}. The reservoir parameters of the cavity array are kept the same as Fig.~\ref{fig:spectrum_CCA}.}
		\label{fig:dressing_CCA}
	\end{figure}
	\subsubsection{Polariton behavior}
	Having looked at the normal mode spectrum, we now turn our attention to the behavior of the polaritonic excitations at different values of the coupling $\theta_{\mr{C/CA}}$. Towards that, we first note that using the Bogoliubov transformation $T$ that defined the operators $\hat{\zeta}_i$ in terms of the EM field and dipole operators given in Eq.~\eqref{eq:PolaritonDefinition} we can write 
	\begin{align}
		\hat{\zeta}_{i}= & \sum_{j=1}^{N_{d}}\left(T_{ij}\hat{a}_{j}+T_{i\left(j+N\right)}\hat{a}_{j}^{\dagger}\right) \label{eq:PolaritonExpansion}\\
		& +\sum_{j=N_{d}+1}^{N_{\mr{tot}}}\left(T_{ij}\hat{R}_{j}+T_{i\left(j+N\right)}\hat{R}_{j}^{\dagger}\right) \nonumber,
	\end{align}
	for the $i^{\mr{th}}$ polariton mode operator in a setting with $N_\mrd$ dipoles and $N$ EM reservoir modes. A clear way to quantify the dressing of light and matter degrees of freedom in a given polariton mode is by looking at the relative weights of the light and matter coefficients given by the elements of the Bogoliubov transformation matrix $T$ appearing in Eq.~\eqref{eq:PolaritonExpansion} \cite{PhysRevLett.112.016401}. In particular we will examine the normalized matter component associated with the $i^{th}$ polariton which is given as
	\begin{align}
		\chi_{i}= & \frac{\sum_{j=1}^{N_{d}}\left\{ \left|T_{ij}\right|^{2}+\left|T_{i\left(j+N\right)}\right|^{2}\right\} }{\sum_{j=1}^{N_{\mr{tot}}}\left\{ \left|T_{ij}\right|^{2}+\left|T_{i\left(j+N\right)}\right|^{2}\right\} }.
	\end{align}
	It is easy to see that $\chi_i$ will become $1$ for polariton modes which are purely composed of matter and $0$ for modes which are purely radiation, while intermediate values mean that the mode consists of both light and matter components. 
	\begin{figure}
		\begin{centering}
			\begin{overpic}[width=0.8\columnwidth]{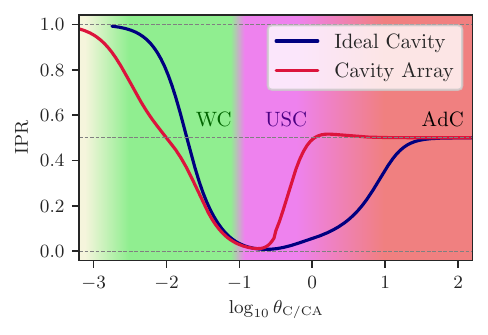}
			\end{overpic}
		\end{centering}
		\caption{Inverse participation ratio (IPR) of the matter degree of freedom over polaritonic modes as a function of the coupling parameter $\theta_{\mr{C}}$ and $\theta_{\mr{CA}}$ for a dipole coupled to an ideal cavity and cavity array reservoirs respectively.}
		\label{fig:IPR}
	\end{figure}
	\begin{figure*}
		\begin{centering}
			\subfloat{\begin{overpic}[width=0.32\linewidth]{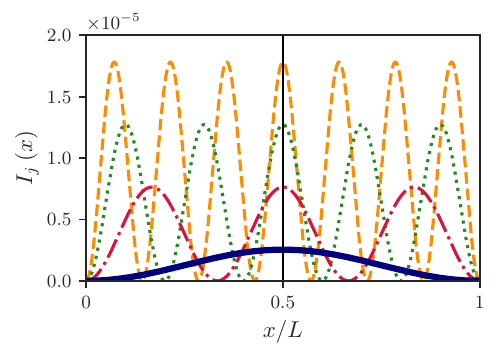}
					\put(25,55){\textbf{(a)}}
					\put(68,58){WC}
				\end{overpic}
			}
			\subfloat{\begin{overpic}[width=0.32\linewidth]{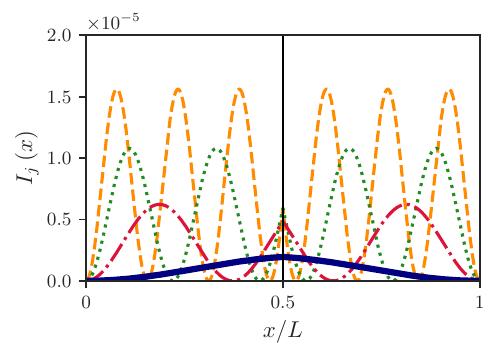}
					\put(25,55){\textbf{(b)}}
					\put(68,58){USC}
				\end{overpic}
			}
			\subfloat{\begin{overpic}[width=0.32\linewidth]{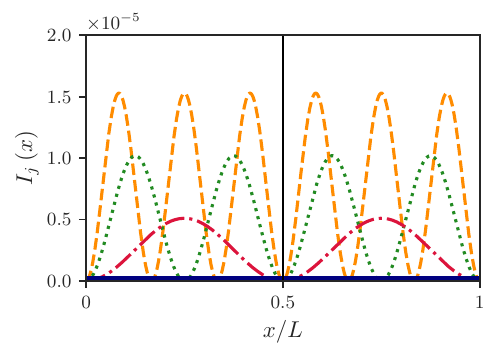}
					\put(25,55){\textbf{(c)}}
					\put(68,58){AdC}
				\end{overpic}
			}
		\end{centering}
		\caption{Field intensity in the single excitation state of the $j^{\mr{th}}$ (odd $j$) polariton mode for a single dipole of frequency $\omega_s = 500 \omega_c$ coupled to an ideal cavity with strengths $\log_{10} \theta_{\mr{C}} = (-3,-0.1,2)$ in (a,b,c) respectively. The solid blue, dash dotted red, dotted green, and dashed orange lines correspond to $j=1,3,5,$ and $7$ respectively and the vertical solid line indicates the location of the oscillator.}
		\label{fig:Polariton_field}
	\end{figure*}
	\begin{figure*}
		\begin{centering}
			\subfloat{\begin{overpic}[width=0.32\linewidth]{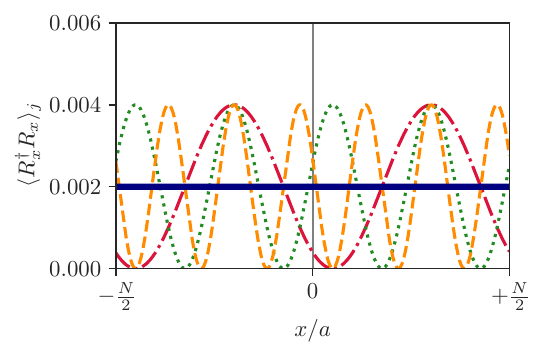}
					\put(25,55){\textbf{(a)}}
					\put(68,55){WC}
				\end{overpic}
			}
			\subfloat{\begin{overpic}[width=0.32\linewidth]{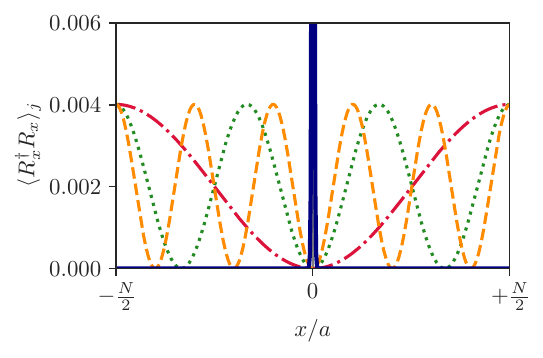}
					\put(25,55){\textbf{(b)}}
					\put(68,55){USC}
				\end{overpic}
			}
			\subfloat{\begin{overpic}[width=0.32\linewidth]{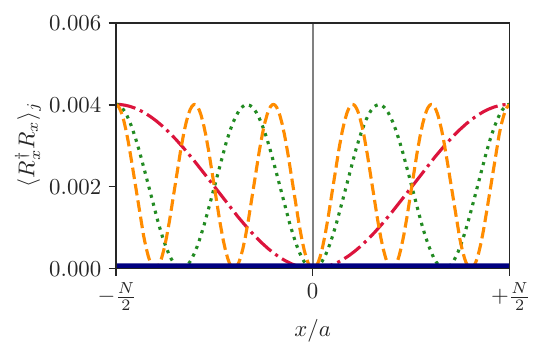}
					\put(25,55){\textbf{(c)}}
					\put(68,55){AdC}
				\end{overpic}
			}
		\end{centering}
		\caption{Photon number in the single excitation state of the $j^{\mr{th}}$ (odd $j$) polariton mode for a single dipole coupled to a cavity array with strengths $\log_{10} \theta_{\mr{CA}} = (-3,-0.1,5)$ in (a,b,c) respectively. The solid blue, dash dotted red, dotted green, and dashed orange correspond to $j=1,3,5,$ and $7$ respectively. The reservoir parameters of the cavity array are kept the same as Fig.~\ref{fig:spectrum_CCA}. Vertical solid line indicates the location of the oscillator.} 
		\label{fig:Polariton_field_CCA}
	\end{figure*}
	
	In Fig.~\ref{fig:dressing} (a) and (b) we plot the matter components for frequencies around the dipole frequency and low normal mode frequencies respectively for the case of a dipole placed in the cavity EM reservoir in different coupling regimes. In the WC regime we see that there is a distribution of the matter wave component between polaritons of different frequencies near the resonant frequency but low energy polaritons have a rather small matter wave component. In contrast, as the coupling is further increased to the USC regime the matter component becomes concentrated in the low energy polariton modes with a peak at the lowest energy mode. This shifting of the peak of the matter component of polariton is further sharpened in the AdC regime and eventually in the asymptotic limit of infinite coupling, all of the matter component is in the lowest energy polariton mode. This is a key signature of asymptotic decoupling \cite{PhysRevLett.112.016401} of the light and matter degrees of freedom and will be used later to understand the system dynamics in this regime. The matter component behavior for the cavity array reservoir shown in Fig.~\ref{fig:dressing_CCA}, as in the case of the spectrum, is largely similar to the cavity. The chief difference is in the USC and AdC regimes, where for the same value of $\theta_{\mr{CA}} = \theta_{\mr{C}}$ the matter component in the cavity array case, as shown in Fig.~\ref{fig:dressing_CCA}(b), is more sharply peaked at the lowest energy polariton mode than in the cavity case. In this sense the asymptotic decoupling is established at even smaller values of $\theta_{\mr{CA}}$ in the cavity array case. Note that in Figs.~\ref{fig:dressing} and ~\ref{fig:dressing_CCA} we only plot the odd $j$ polariton modes as the even ones are uncoupled reservoir modes. 
	
	The behavior of the polaritonic operators and their dressing properties as the coupling is varied can also be summarized by parameterizing the delocalization of the matter degree of freedom in the polariton space. To do this, we write the dipole operators in terms of the polariton operators using the inverse of the Bogoliubov transformation $A = T^{-1}$ as
	\begin{align}
		\hat{a}_{i}= & \sum_{n=1}^{N_{\mr{tot}}}\left\{ A_{in}\hat{\zeta}_{n}+A_{i\left(n+N_{\mr{tot}}\right)}\hat{\zeta}_{n}^{\dagger}\right\} \label{eq:MatterExpansion}. 
	\end{align}
	The coefficients $A_{ij}$ can be used to define a normalized inverse participation ratio (IPR) for the $i^{th}$ dipole as follows
	\begin{align}
		P_{i}= & \frac{\sum_{n=1}^{N_{\mr{tot}}}\left\{ \vert A _{in}\vert^{4}+\vert A_{i\left(n+N_{\mr{tot}}\right)}\vert^{4}\right\} }{\sum_{n=1}^{N_{\mr{tot}}}\left\{ \vert A_{in}\vert^{2}+\vert A_{i\left(n+N_{\mr{tot}}\right)}\vert^{2}\right\} ^{2}}.
	\end{align}
	This quantity, like $\chi_i$, is bounded between zero and one, where zero implies delocalisation over all available polariton operators and one means that the matter degree of freedom is confined to one polariton operator. The IPR for a single dipole coupled to ideal cavity and cavity array reservoirs are shown in Fig.~\ref{fig:IPR}. The different regimes of light-matter interaction can then be neatly characterized by the IPR. In the very weak coupling regime ($\theta_{\mr{C/CA}} \ll 10^{-1}$), the light-matter dressing is negligible and the Bogoliubov transformation is almost equivalent to an identity matrix and we find $A_{n} \approx \delta_{n,n_{\mr{res}}}A_{n_{\mr{res}}}$ (dropping the index $i$ as there is only one dipole). This shows that the matter mode gets approximately mapped to a single `resonant' polariton mode indicating the absence of dressing leading to the IPR to go to $1$ for both the cavity and cavity array reservoir as $\theta_{\mr{C/CA}} \rightarrow 0$. As $\theta_{\mr{C/CA}}$ increases, the delocalization around the resonant polariton increases reducing the IPR. This delocalization and lowering of the IPR is a characteristic of the WC regime and continues up to the transition to USC ($\theta_{\mr{C}}\sim 10^{-1}$), where we start seeing a peak around the lowest frequency polariton mode in plots for the matter component (Figs.~\ref{fig:dressing} and ~\ref{fig:dressing_CCA}). This localization around the low frequency modes leads to an increase in the IPR, which goes asymptotically to $0.5$ deep in the AdC regime where we find $A_{n} \approx  \delta_{n,1}/\sqrt{2} + \delta_{n,N_{\mr{tot}}+1}/\sqrt{2}$ i.e. the matter mode is entirely concentrated in the lowest energy polariton. Note that Fig.~\ref{fig:IPR} also shows that the IPR approaches its asymptotic value of $0.5$ faster for the cavity array reservoir as compared to the ideal cavity. This is in agreement with the earlier comment that asymptotic decoupling happens at smaller values of $\theta_{\mr{CA}}$ for the cavity array. It is to be noted that the way we have defined IPR acknowledges the fact that in the regimes with large light-matter coupling the matter operator has to be written in terms of both the annihilation and creation operators of the polaritons. While in the very weak coupling regimes the contribution of the creation operators is small $A_{i(n+N_{\mr{tot}})} \approx 0$, in the USC and AdC regimes $A_{i(n+N_{\mr{tot}})}$ becomes significant ultimately becoming as large as $A_{in}$. This indicates that in the USC and AdC regimes the quantum vacuum of the polaritons are very different from the original light and matter degrees of freedom \cite{frisk_kockum_ultrastrong_2019}.
	
	\begin{figure}
		\begin{centering}
			\subfloat{\begin{overpic}[abs,width=0.8\columnwidth]{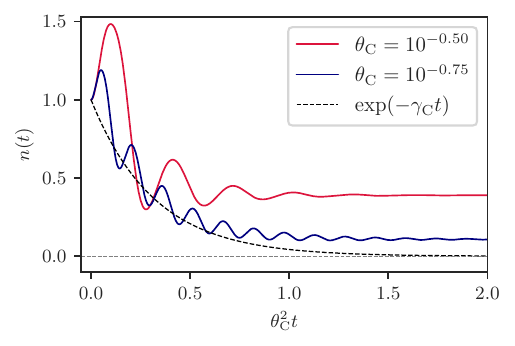}
					\put(70,90){\textcolor{black}{\textbf{(a)}}}
				\end{overpic}
			}\\
			\subfloat{\begin{overpic}[abs,width=0.8\columnwidth]{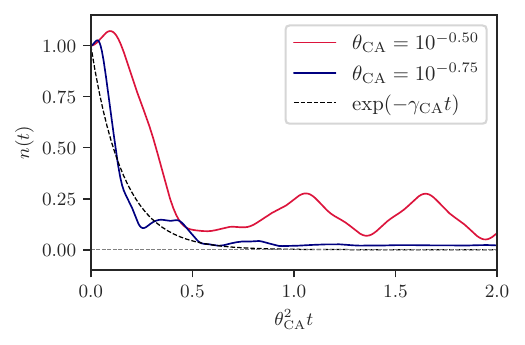}
					\put(70,90){\textcolor{black}{\textbf{(b)}}}
				\end{overpic}
			}\\
		\end{centering}
		\caption{Population dynamics of a single dipole oscillator coupled to (a) cavity and (b) cavity array reservoir in the ultrastrong coupling regime (USC) with different coupling strengths. The associated Markovian decay dynamics at the rate $\gamma_{\mr{C/CA}}$ is shown with dashed black lines in the plot. System and reservoir parameters for the cavity and cavity array are kept the same as Figs.~\ref{fig:classification} and ~\ref{fig:spectrum_CCA} respectively. We include $N=4000$ and $N=500$ modes for the cavity and cavity array reservoir respectively in the numerical calculations.
		}
		\label{fig:1e_USCweak}
	\end{figure}
	A clear way to visualize this change in the vacuum as the coupling is made stronger and understand the asymptotic decoupling is by looking at the position space properties corresponding to the polariton Fock states $\hat{\zeta}_n^{\dagger} \tilde{\ket{0}}$ with a single excitation, where the $\tilde{\ket{0}}$ state represents the dressed vacuum i.e. $\hat{\zeta}_n \tilde{\ket{0}} = 0$ \cite{PhysRevLett.112.016401}. For the cavity reservoir, we consider the intensity of the electric field inside the cavity in a state with a single polariton $\zetaop_j$ which is given by
	\begin{align}
		I_j(x) = \tilde{\bra{0}}\zetaop_j \Eop^{-}(x) \Eop^{+}(x) \zetaop_j^{\dagger} \tilde{\ket{0}}, \label{eq:polfockI_cav}
	\end{align}
	with the operators $\Eop^{\pm}(x)$ defined in Eq.~\eqref{eq:ReservoirEfield}. $I_j(x)$ has been plotted in Fig.~\ref{fig:Polariton_field} for the low frequency odd polaritons (even ones are uncoupled reservoir modes) in the very weak limit of the WC, USC, and AdC coupling regimes for a single dipole at the center of the cavity. In the very weak coupling limit, the field modes are undressed and correspond to the standing waves inside the cavity. Since we have considered only odd polaritons, anti-nodes are expected at the position of the dipole. As the coupling is increased to the USC and AdC regimes, we see a gradual decrease in the electric field amplitude and eventually a node of the electric field emerges for all polaritons except for the lowest energy one at the position of the dipole \cite{PhysRevLett.112.016401}. Thus the electric field is localized away from the position of the dipole and the cavity is effectively split into parts. Moreover the electric field for the lowest energy polariton $\zetaop_1$, which is essentially the field of  fundamental cavity mode in the very weak coupling, eventually vanishes in the $\theta_{\mr{C}} \rightarrow \infty$ asymptotic limit and the polariton becomes completely matter-like. This is precisely the asymptotic decoupling between light and matter that arises from the important role of the $\bm{\hat{A}}^2$ term in the Hamiltonian as pointed out in detail in \cite{PhysRevLett.112.016401}. For the cavity array reservoir, the position space observable of interest is the on-site photonic population introduced in Eq.~\eqref{eq:CavityArrayPhotNumPosn} in the polariton Fock state $\zetaop_j^{\dagger} \tilde{\ket{0}}$ given as 
	\begin{align}
		\avg{\hat{R}_{x}^{\dagger}\hat{R}_{x}}_j= & \tilde{\bra{0}} \hat{\zeta}_{j}\hat{R}_{x}^{\dagger}\hat{R}_{x}\hat{\zeta}_{j}^{\dagger}\tilde{\ket{0}}. \label{eq:polfockpop_CCA}
	\end{align}
	In Fig.~\ref{fig:Polariton_field_CCA} we plot $\avg{\hat{R}_{x}^{\dagger}\hat{R}_{x}}_j$ for the odd polaritons in the three different coupling regimes for a dipole placed at the central cavity of the array. In the very weak coupling limit these polaritons (except for the lowest energy polariton with wave number $k_n=0$) correspond to the even superposition of the traveling waves moving in opposite directions ($k_n= \pm n 2 \pi/L$) with degenerate frequency and inside the cavity array (see Appendix \ref{app:C}). As the coupling strength is increased, we see the same behavior that is observed in the cavity case and the photonic population goes to zero at the position of the dipole for all but the lowest energy polariton mode showing that asymptotic decoupling is a general feature of strong light-matter coupling.
	regime. 
	
	\subsection{Dynamics in the Strong Interaction Regimes}
	In this subsection, we present the numerical results for the dynamics of a single oscillator dipole and two dipole oscillators strongly coupled to its electromagnetic environment in the USC and AdC regimes.
	\subsubsection{Single Oscillator Dynamics}
\begin{figure}
		\begin{centering}
			\subfloat{\begin{overpic}[abs,width=0.85\columnwidth]{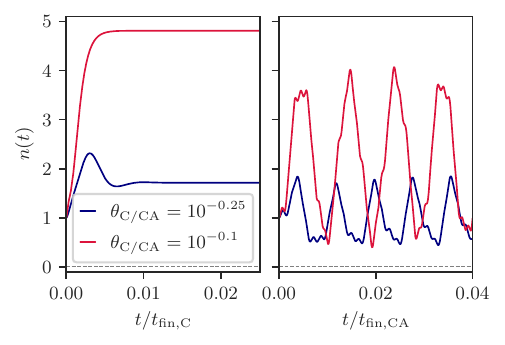}
					\put(90,113){\textbf{(a)}}
					\put(176,113){\textbf{(b)}}
				\end{overpic}
			}
		\end{centering}
		\caption{Population dynamics of a single dipole oscillator coupled to (a)  an ideal cavity and (b) a cavity array for two different coupling strengths deep in the USC regime. System and reservoir parameters are kept the same as in Fig.~\ref{fig:1e_USCweak}.}
		\label{fig:USC}
	\end{figure}

	Focusing first on the single oscillator case in the USC regime, we find two qualitatively different kinds of dynamics depending on the value of the coupling. For coupling regimes contiguous with the WC regime with $\theta_{\mr{C/CA}} \gtrsim 10^{-1}$, the behavior is as shown in Figs.~\ref{fig:1e_USCweak} (a) and (b) for the cavity and cavity array respectively. Both these cases show the dynamics for the dipole initially in a Fock state with a single excitation. The dynamics of the oscillator excitation has an overall decay with approximately same rate as the Markovian decay $\gamma_{\mr{C/CA}}$ discussed in Sec.~\ref{sec:WeakIntMarkov} which scales as $\gamma_{\mr{C/CA}} \sim \theta_{\mr{C/CA}}^2$. On top of this decay, we see large amplitude oscillations as well as regions where the oscillator has more excitations than in the initial state \footnote{Note that these oscillations with much smaller amplitude and large frequency were present even in the WC regime and in Sec.~\ref{sec:WeakIntMarkov} we had to coarse grain over them to get agreement with Lindblad master equation.}. These effects owe their origin to the counter-rotating terms that become important in the USC regime. In addition to this we also see that unlike in the WC regime, for the cavity reservoir [Fig.~\ref{fig:1e_USCweak} (a)] we clearly see a steady state with finite excitation of the oscillator emerge. In contrast for the cavity array case [Fig.~\ref{fig:1e_USCweak} (b)], even though there is decay, there is no steady state as the oscillator population continues to oscillate. As the coupling is increased further in the USC regime, as we show in Fig.~\ref{fig:USC}, for the cavity reservoir case the oscillations become overdamped and a steady state with large value of excitation $n(t) = \avg{\adop \aop}$ emerges. In contrast, for the cavity array case, while the average excitation is of the same order of magnitude as the cavity, there are undamped oscillations. It is to be noted that these plots show the dynamics over timescales smaller than the unavoidable finite-size revival timescales $t_{\mr{fin},C/CA}$.
	regime.       
	\begin{figure}
		\begin{centering}
			\subfloat{\begin{overpic}[abs,width=0.8\columnwidth]{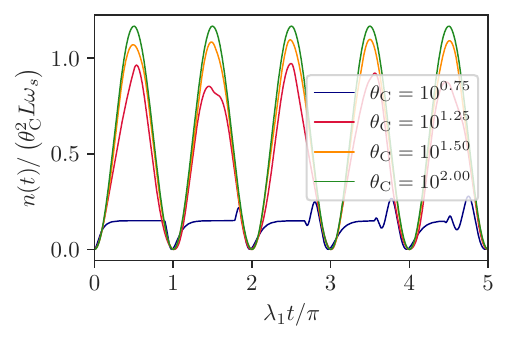}
					\put(64,90){\textcolor{black}{\textbf{(a)}}}
				\end{overpic}
			}\\
			\subfloat{\begin{overpic}[abs,width=0.8\columnwidth]{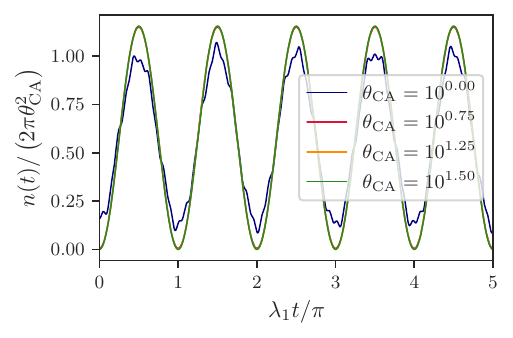}
					\put(64,90){\textcolor{black}{\textbf{(b)}}}
				\end{overpic}
			}\\
		\end{centering}
		\caption{Population dynamics of single dipole oscillator coupled to (a) cavity and (b) cavity array reservoir in the asymptotically decoupled regime shown at different coupling strengths. System and reservoir parameters are kept the same as in Fig.~\ref{fig:1e_USCweak}.
		}
		\label{fig:1e_AdC}
	\end{figure}
	
	The dynamics in the AdC regime is presented in Fig.~\ref{fig:1e_AdC} (a) and (b) for the cavity and cavity array case respectively. In both cases the dynamics is oscillatory and as the coupling strength is increased a single sinusoidal oscillation at periodicity given by $1/(2 \lambda_1)$ emerges. Recall that $\lambda_1$ is the smallest polaritonic mode's frequency which goes to zero asymptotically in the limit of large $\theta_{\mr{C/CA}}$. Secondly, as the interaction strength is increased the amplitude of the oscillation scales as $\sim \theta_{\mr{C}}^2 L$ in the cavity case and $\sim \theta_{\mr{CA}}^2$ in the cavity array. Finally, in both Fig.~\ref{fig:1e_AdC} (a) and (b) we have plotted the dynamics in the USC regime (blue lines) contiguous to the AdC regime for comparison. But unlike Fig.~\ref{fig:USC}, we have plotted the dynamics over multiple finite size recurrence times. This illustrates the fact that the oscillatory dynamics in the AdC regime emerges from the participation of a polaritonic mode with spatial delocalization extending over the entire reservoir. Moreover, while there is a clear qualitative transition in the dynamics from the USC to AdC for the cavity reservoir, the USC regime dynamics for the cavity array is qualitatively similar to the AdC regime (oscillatory). In this sense the USC regime for the cavity array appears to be a smooth transition region from the irreversible WC regime dynamics to oscillatory AdC regime dynamics. 
	\begin{figure}
		\begin{centering}
			\subfloat{\begin{overpic}[abs,width=0.8\columnwidth]{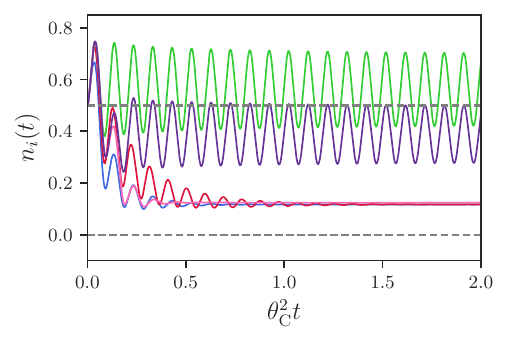}
					\put(45,45){\textcolor{black}{\textbf{(a)}}}
				\end{overpic}
			}\\
			\subfloat{\begin{overpic}[abs,width=0.8\columnwidth]{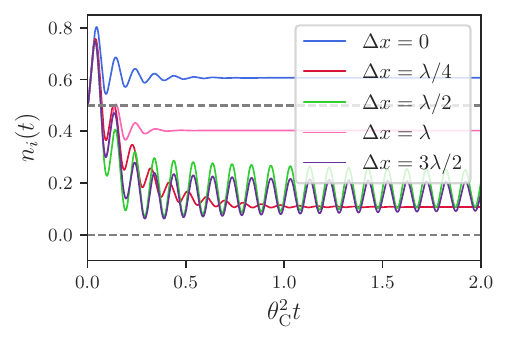}
					\put(45,45){\textcolor{black}{\textbf{(b)}}}
				\end{overpic}
			}
		\end{centering}
		\caption{Excitation number dynamics of two dipoles
			strongly coupled ($\theta_\mr{C} = 10^{-0.75}$) to a cavity
			plotted for different separations $\Delta x$ with the initial state $\ket{\psi_+}$ (a) and $\ket{\psi_-}$ (b) with $\ket{\psi_{\pm}} = \left(\ket{10} \pm \ket{01} \right)/\sqrt{2}$. System and reservoir parameters are kept the same as in Fig.~\ref{fig:1e_USCweak}.}
		\label{fig:SC_2e}
	\end{figure}
	\begin{figure}
		\begin{centering}
			\subfloat{\begin{overpic}[abs,width=0.8\columnwidth]{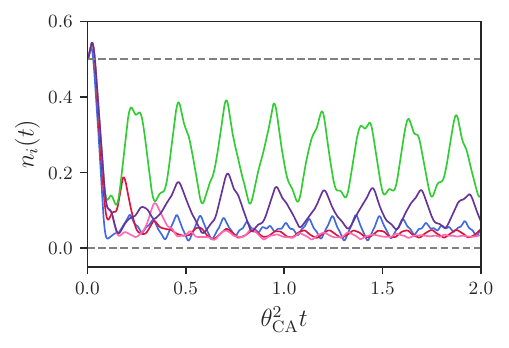}
					\put(70,80){\textcolor{black}{\textbf{(a)}}}
				\end{overpic}
			}\\
			\subfloat{\begin{overpic}[abs,width=0.8\columnwidth]{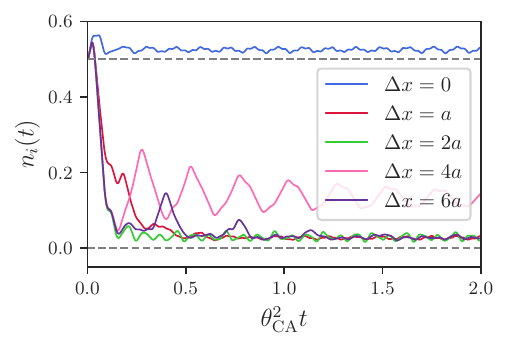}
					\put(70,80){\textcolor{black}{\textbf{(b)}}}
				\end{overpic}
			}
		\end{centering}
		\caption{Excitation number dynamics of two dipoles
			strongly coupled ($\theta_\mr{CA} = 10^{-0.5}$) to a cavity array plotted for different separations $\Delta x$ with the initial state $\ket{\psi_+}$ (a) and $\ket{\psi_-}$ (b). System and reservoir parameters are kept the same as in Fig.~\ref{fig:1e_USCweak}.}
		\label{fig:SC_2e_CCA}
	\end{figure}
	\begin{figure}
		\begin{centering}
			\begin{overpic}[abs,width=0.8\columnwidth]{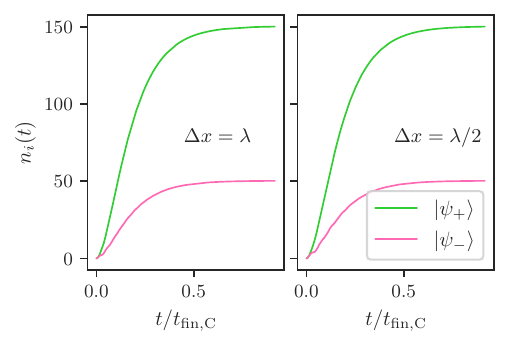}
				\put(90,110){\textcolor{black}{\textbf{(a)}}}
				\put(170,110){\textcolor{black}{\textbf{(b)}}}
			\end{overpic}
		\end{centering}
		\caption{Excitation number dynamics of two dipoles strongly coupled to an ideal cavity and placed with separations $\Delta x = \lambda,\lambda/2$ in (a,b) respectively. The coupling is deep in the USC regime with $\theta_{\mr{C}} = 10^{-0.1}$  and the initial states are taken as $\ket{\psi_\pm}$. Other parameters are kept the same as in Fig.~\ref{fig:1e_USCweak}.}
		\label{fig:USC_deep_2e_cav}
	\end{figure}
	\subsubsection{Collective Dynamics - Strong Interaction Regimes}
	Having presented the dynamics of a single dipole coupled strongly to its environment, let us now examine how the rich collective effects that are observed in the WC regime (discussed in Sec.~\ref{sec:WeakIntMarkov}) are modified in the strong interaction regime. In particular, we consider the dipole population dynamics for two oscillator dipoles coupled to the reservoir in the strong coupling regimes. We find that this is enough to understand the general collective behavior and will briefly comment on the many oscillator case towards the end.
	
	Focusing first on the USC regime, as in the single dipole case we find two qualitatively different dynamical behavior. For $\theta_{\mr{C/CA}} \gtrsim 10^{-1}$, the two dipole oscillator populations for the cavity and cavity array are shown in Figs.~\ref{fig:SC_2e} and ~\ref{fig:SC_2e_CCA} respectively. In this regime the observed dynamics has many qualitative similarities with the collective dynamics of two dipole oscillators in the WC regime displayed in Figs.~\ref{fig:2e_cav} and \ref{fig:2e_CCA} but there are also some important differences. In particular, we again see that for dipoles initially in a superradiant state [Fig.~\ref{fig:SC_2e}(a) and Fig.~\ref{fig:SC_2e_CCA}(a)], a fast decay leads to a steady state with finite excitation (which is not the ground state of the dipole oscillators) for integer wavelength separations. The rate of decay can again be modified by changing the separation between the dipole oscillators like in the WC regime. At half wavelength separation, we obtain subradiant dynamics in the sense that there is no decay, however we see large oscillations about the subradiant population. We emphasize again that these oscillations are of much larger amplitude from those we coarse-grain over in the WC regime. Moreover if we indeed coarse-grain over these oscillations, for instance in for the $\Delta x = \lambda$ separation case in the cavity, we obtain a steady state larger than the initial average population of the dipoles which is half. This behavior is also more clearly demonstrated for the plots with a subradiant initial state, shown in Figs.~\ref{fig:SC_2e}(b) and \ref{fig:SC_2e_CCA}(b), where the subradiant population stabilises to a steady population greater than half for the zero separation case. 
	\begin{figure}
		\begin{centering}
			\subfloat{\begin{overpic}[abs,width=0.8\columnwidth]{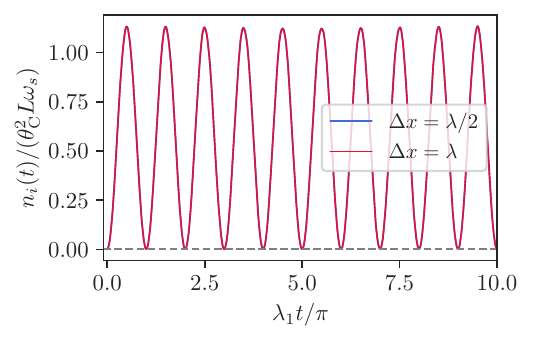}
					\put(150,50){\textcolor{black}{\textbf{(a)}}}
				\end{overpic}
			}\\
			\subfloat{\begin{overpic}[abs,width=0.8\columnwidth]{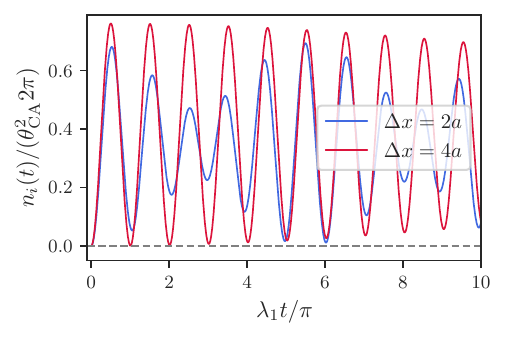}
					\put(150,50){\textcolor{black}{\textbf{(b)}}}
				\end{overpic}
			}\\
		\end{centering}
		\caption{Population dynamics of two dipoles placed (a) symmetrically about the center of a cavity and (b) at two different sites of a cavity array with separation $\Delta x$ in the AdC regime with $\theta_{\mr{C/CA}} = 10^{1.5}$. The two emitters are initially in a subradiant state $\vert \psi_{-} \rangle$. Other parameters are kept the same as in Fig.~\ref{fig:1e_USCweak}.}
		\label{fig:AdC_2e}
	\end{figure}
	
	Moving to larger couplings in the USC regime, as in the single dipole case, we find that for the cavity reservoir the oscillatory dynamics is replaced by overdamped dynamics to a steady state with higher excitation number as shown in Fig.~\ref{fig:USC_deep_2e_cav}. Moreover, the dynamics becomes independent of the dipole separation $\Delta x$. We also find that this behavior also persists to arrays with multiple dipole oscillators (see Fig.~\ref{fig:10e_pop_USC} in Appendix \ref{app:C}). In fact, this separation independence of the dynamics is even more clear in the AdC regime, shown in Fig.~\ref{fig:AdC_2e} (a), where the excitation number becomes sinusoidal with frequency given, as in the single dipole case, by twice the lowest polariton mode frequency $2 \lambda_1$ and amplitude scaling as $\theta_{\mr{C}}^2 L$. In contrast, we find that for the cavity array the dynamics continues to be oscillatory and there is separation dependence. Since this behavior is qualitatively similar in the USC and AdC regimes, we display the same for the AdC regime in Fig.~\ref{fig:AdC_2e} (b). As we can see there, in addition to the oscillatory feature at the frequency $2 \lambda_1$, there is an additional slower oscillatory envelope whose frequency is not the same for different separations of the dipole oscillators. In the next subsection we will understand the origin of this envelope frequency.
	
	\subsection{Brief Analysis - Strong Coupling Regime Results}
	
	In the previous sub-section we have outlined, using the results of exact numerical simulation, the rich variety of dynamics that opens up for a system of dipoles strongly coupled to a cavity or cavity array EM reservoir. In this section we would like to provide a heuristic understanding of the observed dynamics beginning with the AdC regime where the dynamics simplifies greatly.
	
	As explained in the previous sub-sections, deep in the AdC regime the matter content of only the lowest energy polariton is non-zero and hence we can write the annihilation operator of the single coupled dipole as
	\begin{align*}
		\hat{a} &\approx A_{11}\hat{\zeta}_{1}+A_{1\left(N_{\mr{tot}}+1\right)}\hat{\zeta}_{1}^{\dagger},
	\end{align*}
	with $\{A_{11},A_{1\left(N_{\mr{tot}}+1\right)}\}\rightarrow 1/\sqrt{2}$ in the limit of $\theta_{\mr{C/CA}}\rightarrow \infty$. Using the time evolution of the polariton mode as $\zetaop_i = \zetaop_i e^{-i \lambda_i t}$, the time dependence of the population dynamics can be written as
	\begin{align*}
		&\avg{\hat{a}^{\dagger}a\left(t\right)}=  \left|A_{11}\right|^{2}\avg{\hat{\zeta}_{1}^{\dagger}\hat{\zeta}_{1}}+A_{1\left(N_{\mr{tot}}+1\right)}A_{11}^{*}\avg{\hat{\zeta}_{1}^{\dagger}\hat{\zeta}_{1}^{\dagger}}e^{+i2\lambda_{1}t}\\
		& +A_{11}A_{1\left(N_{\mr{tot}}+1\right)}^{*}\avg{\hat{\zeta}_{1}\hat{\zeta}_{1}}e^{-i2\lambda_{1}t}+\left|A_{1\left(N_{\mr{tot}}+1\right)}\right|^{2}\avg{\hat{\zeta}_{1}\hat{\zeta}_{1}^{\dagger}}.
	\end{align*}
	Thus, the only time-dependence comes from the counter-rotating correlations which are oscillating with a frequency of $2\lambda_1$. This is exactly the behavior we see in the AdC population dynamics pictured in Figs.~\ref{fig:1e_AdC} and \ref{fig:AdC_2e}. This oscillatory dynamics is essentially induced by the strong dressing which squeezes the matter degrees of freedom into the low frequency polariton mode. The other AdC polaritons modes correspond purely to the radiation degrees of freedom. The observed dynamics is thus a manifestation of this separation or asymptotic decoupling of the light and matter degrees of freedom matter at very large system-reservoir interaction strengths \cite{PhysRevLett.112.016401,Ashida21,Ashida22}. 
	\begin{figure}
		\begin{centering}
			\subfloat{\begin{overpic}[abs, width=0.9\columnwidth]{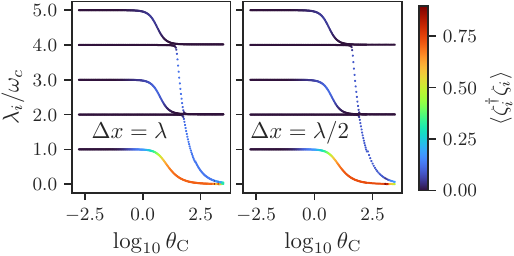}
					\put(80,100){\textbf{(a)}}
					\put(150,100){\textbf{(b)}}
			\end{overpic}}\\
			\subfloat{\begin{overpic}[abs,width=0.9\columnwidth]{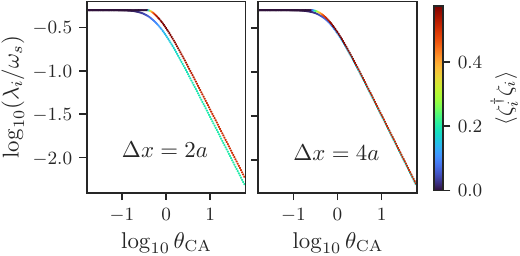}
					\put(80,90){\textbf{(c)}}
					\put(150,90){\textbf{(d)}}
			\end{overpic}}
		\end{centering}
		\caption{Low energy polariton frequencies (lines) and their corresponding normalized populations (line colour with legends on the right) for two dipoles coupled strongly to a cavity (a,b) and cavity array (c,d) reservoirs with different separations $\Delta x$ with other parameters same as in Fig.~\ref{fig:AdC_2e}.}
		\label{fig:Beats_discussion}
	\end{figure}
	\begin{figure}
		\begin{centering}
			\subfloat{\begin{overpic}[abs,width=0.8\columnwidth]{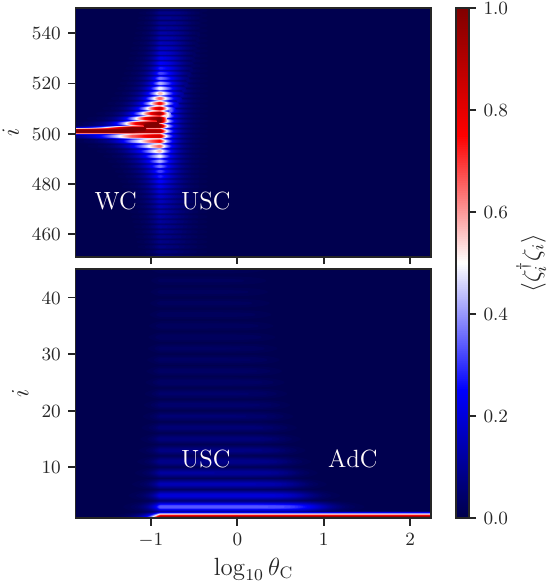}
					\put(100,160){\textcolor{white}{\textbf{(a)}}}
					\put(100,80){\textcolor{white}{\textbf{(b)}}}
				\end{overpic}
			}
		\end{centering}
		\caption{Population distribution in the polariton space as a function of coupling parameter $\theta_\mr{C}$ for a single oscillator coupled to an ideal cavity with the initial state $\vert 1 \rangle \otimes \vert 0 \rangle_R$. The top and the bottom panel show the polaritons modes near the resonant energy level $\omega_s = 500\omega_c$ and low frequencies respectively. Other parameters are kept the same as in Fig.~\ref{fig:1e_USCweak}.}
		\label{fig:polariton_pop}
	\end{figure}
	
	Another important feature of the AdC regime is the absence of separation dependent collective effects for the cavity, whereas dynamics for the cavity array reservoir shows separation dependent beats even in the AdC regime [see Fig.~\ref{fig:AdC_2e}(a,b)]. The reason for this absence of the collective effects may be understood by looking at the polariton populations of the low energy polariton modes in the two cases since they control the dynamics in the AdC regime. The variation of normal mode frequencies with the coupling parameter has been shown in Fig.~\ref{fig:Beats_discussion} along with the respective normalised polariton populations in color. A key difference to be be noticed in this plot is that in all cases there are two polariton modes that asymptotically tend to zero frequency as the coupling is increased. Focusing first on the cavity case shown in Fig.~\ref{fig:Beats_discussion} (a,b) for two dipole oscillator separations $\Delta x = \lambda, \lambda/2$ respectively, we see first that the frequency of the lowest energy polariton is completely independent of the separation $\Delta x$. Even though the first excited polariton mode (that also goes to zero asymptotically) indeed has some dependence on $\Delta x$, the population of this mode is completely negligible until very large values of $\theta_{\mr{C}}$ at which point it is almost degenerate with the lowest energy polartiton mode. As a result the dynamics for $\Delta x = \lambda, \lambda/2$ are essentially the same in the large coupling AdC regime. Coming to the contrasting behavior in the cavity array reservoir case, in Fig.~\ref{fig:Beats_discussion} (c,d) we show the behavior of the two lowest polariton modes for the cavity array reservoir with dipole oscillators placed with separations of $2a$ and $4a$. While the lowest energy mode is again separation independent, the energy difference between the lowest and the second lowest energy mode (which also goes to zero at large $\theta_{\mr{CA}}$), $\lambda_2-\lambda_1$, is separation dependent. Moreover, the population of these two low energy modes are of the same order of magnitude in much of the AdC regime. Thus, in addition to the oscillation at the separation independent frequency of the lowest energy polariton mode $2\lambda_1$, the dynamics also shows periodicity at a separation dependent beat frequency $\lambda_2-\lambda_1$ for cavity array case. Finally, a picture of the oscillatory dynamics in the AdC regime as an exchange of excitation between the dipole oscillator and the cavity to which it is coupled to in the array also emerges from the dynamics of $\avg{\Rop_x^\dagger \Rop_x (t)}$ (see Fig.~\ref{fig:Bath_dyn_CCA} in Appendix \ref{app:C}).
	
\begin{figure}
		\begin{centering}
			\subfloat{\begin{overpic}[abs,width=0.8\columnwidth]{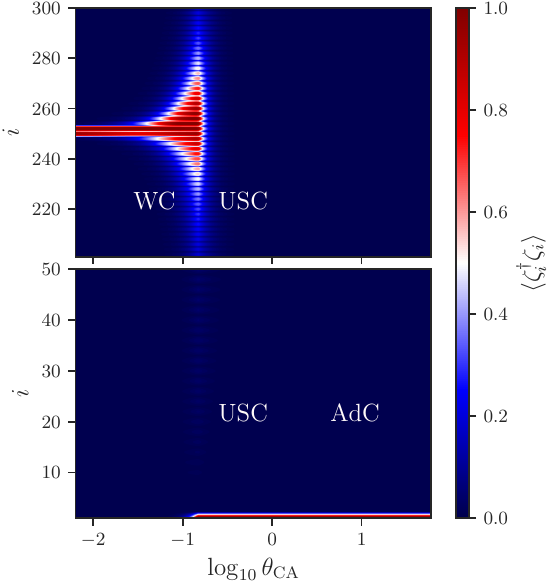}
					\put(100,160){\textcolor{white}{\textbf{(a)}}}
					\put(100,80){\textcolor{white}{\textbf{(b)}}}
				\end{overpic}
			}\\
		\end{centering}
		\caption{Population distribution in the polariton space as a function of coupling parameter $\theta_\mr{CA}$ for a single oscillator coupled to a cavity array with the initial state $\vert 1 \rangle \otimes \vert 0 \rangle_R$. The top and the bottom panel show the polaritons modes near the resonant energy level $\omega_s = \omega_c$ and low frequencies respectively. Other parameters are kept the same as in Fig.~\ref{fig:1e_USCweak}.
		}
		\label{fig:polariton_pop_CCA}
	\end{figure}

	Coming to the dynamics in the USC regime, since this is a non-perturbative regime \cite{Ashida21,Ashida22}, obtaining simple analytical solutions is not straight forward. Nonetheless, we can provide qualitative insights by considering the distribution of polaritonic occupation numbers $\avg{\zetaop_i^\dagger \zetaop_i}$ of the initial state of the system. Note that since the time evolution of the polaritonic operators are simple ($\zetaop_i(t) =  \zetaop_i e^{-i\lambda_i t}$), this distribution is stationary. The contour plots in Figs.~\ref{fig:polariton_pop} and \ref{fig:polariton_pop_CCA} show the polariton occupation number as a function of the coupling strength $\theta_{\mr{C/CA}}$ is increased for the case of a single dipole initially in a Fock state with one excitation coupled to a cavity and cavity array environment (initially in their vacuum states) respectively. The top panel shows the polaritonic occupation number of modes that are near resonance with the dipole oscillator and the bottom panel shows the same for low energy modes. As emphasized in the discussion regarding matter component of the polaritons, comparing Figs.~\ref{fig:polariton_pop} and \ref{fig:polariton_pop_CCA} reinforces the point that the transition from WC to AdC via the USC happens at much smaller coupling strength for the cavity array case in the sense that there is very little change in the distribution in the cavity array case from the middle of the USC region $\theta_{\mr{CA}}\gtrsim 10^{-0.5}$. Moreover, the key feature of an overdamped increase to a steady state with larger excitation of the dipole mode deep in the USC regime pictured in Fig.~\ref{fig:USC} (a) for the cavity reservoir is clearly correlated with the occupation of low energy polaritons. In fact this steady state behavior emerges essentially when the polaritonic occupation number peaks at the lowest normal mode frequency. Comparing Figs.~\ref{fig:polariton_pop} (b) and \ref{fig:polariton_pop_CCA} (b) in this regime, we notice that for the cavity array case there is occupation of only the lowest energy polariton mode i.e. there is no distribution over low energy polariton modes unlike the cavity reservoir. This lack of bandwidth in `polariton' space around the low frequency peak for the cavity array reservoir is directly manifested in the undamped oscillatory dynamics deep in the USC regime and this naturally continues to the sinusoidal dynamics in the AdC regime. Finally, while we have mainly discussed the dynamics of one and two dipoles in this section, the central behavior, namely the collective spatial separation dependent dynamics is not significant in the strong coupling regimes, carries through also to arrays of oscillators as we illustrate with some examples in Appendix \ref{app:C}.
	
	\section{Conclusion}
	\label{sec:conclusion}
	We have studied the exact dynamics of a collection of harmonic oscillator dipoles interacting with two kinds of one-dimensional (1-D) multimode EM reservoirs, namely one representing an ideal cavity and the other a cavity array. The quadratic nature of the Hamiltonian enables an exact numerical solution of the (system) oscillator dipoles and the reservoir modes. We began by presenting the results of the dynamics in the weak coupling (WC) regime in Sec.~\ref{sec:WeakIntMarkov}, showing good agreement with the Markovian Lindblad master equation predictions. In addition, we studied the dynamics of different configuration of arrays of dipoles in 1-D that also showed additional non-Markovian dynamics even at weak coupling. In Sec.~\ref{sec:stronginteraction} we analyzed the strong coupling regimes starting with a scheme to classify the coupling regimes into the ultrastrong coupling (USC) and asymptotic decoupling regimes (AdC). While the USC regime is characterized by strong delocalization of the polariton modes in terms of the bare dipole and reservoir modes, the AdC regime marks the domination of a single polariton mode with the lowest frequency. In terms of oscillator population dynamics, for the cavity reservoir, the weaker part of USC regime is marked by increasing quantitative deviation from the Markovian dynamics, and eventual development of steady states with larger population than in the initial state. In contrast, for the cavity array reservoir the USC regime serves as essentially a transitory regime from the WC to AdC regime. The asymptotic decoupling of light and matter induced by the prominent role of the $\mathbf{\Aop}^2$ term in the AdC regime leads to simple reversible oscillatory dynamics for both the cavity and cavity array reservoir. Coming to the collective dynamics in the strong coupling regime, once in the deep USC and AdC regime, we generally find that the many interesting oscillator separation dependent dynamics (superradiant and subradiant dynamics) seen in the WC are absent. Qualitatively the dynamics resembles single oscillator evolution in some ways except for some subtle separation dependent features in the cavity array case.
	
	While we have focused on $1$-D EM reservoirs in this work, an interesting extension would be to consider EM field configurations in 2 and 3 dimensions. From the point of view of the exact calculations, apart from an increase in the number of modes, this should not lead to any particular difficulties. Such a treatment would help in understanding to what extent some important predictions from Lindblad master equation studies of 3-D vacuum mediated collective dynamics of quantum emitters such as improved light storage via sub-radiant states in 1-D arrays \cite{asenjo-garcia_exponential_2017}, and enhanced optical cross-section with 2-D arrays \cite{PhysRevLett.116.103602}, hold for exact dynamics of oscillators dipoles. Moreover, in this approach one can also study the effect of stronger coupling on such collective phenomena.
	
	\section*{Acknowledgements}
	This work was supported and enabled by a Sabarmati Fellowship of IIT Gandhinagar, Chanakya Fellowship of I-HUB Quantum Technology Foundation, Pune, India (S.G.) and a CSIR-UGC NET Ph.D. fellowship (I.B.). This research was supported in part by the  International Centre for Theoretical Sciences (ICTS) for participating in the program - Periodically and quasi-periodically driven complex systems (code:ICTS/pdcs2023/6) and we acknowledge the participants for their discussions on the early version of this work. BKA acknowledges CRG Grant No.~CRG/2023/003377 from SERB, Government of India. BKA would like to acknowledge funding from the National Mission on Interdisciplinary  Cyber-Physical  Systems (NM-ICPS)  of the Department of Science and Technology,  Govt.~of  India through the I-HUB  Quantum  Technology  Foundation, Pune, India.
	
	\appendix
	\section{Details regarding the numerical method and gauge invariance}
	\setcounter{figure}{0}
	\renewcommand{\thefigure}{A\arabic{figure}}
	\label{app:A}
	\subsection{Spectrum comparison for Coulomb gauge, dipole gauge and dipole interaction Hamiltonians}
\begin{figure}
		\begin{centering}
			\subfloat{\begin{overpic}[abs,width=0.8\columnwidth]{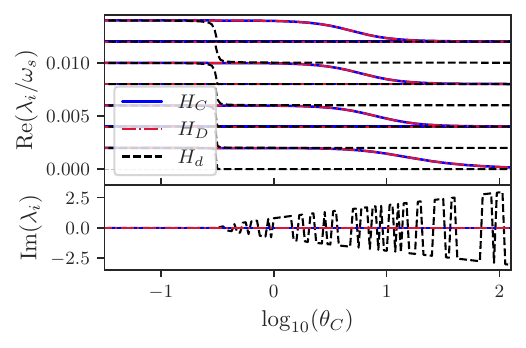}
				\end{overpic}
			}
		\end{centering}
		\caption{Real and imaginary parts of the normal mode spectrum for a single dipole oscillator placed at the center of an ideal cavity as the coupling parameter $\theta_\mr{C}$ is increased for the Coulomb gauge (blue solid line), dipole gauge (red dash-dotted), and quantum optical dipole Hamiltonian (black dashed).}
		\label{fig:spectrum_comparison}
	\end{figure}

	Fig.~\ref{fig:spectrum_comparison} shows the normal mode spectrum for a single dipole oscillator placed at the center of an ideal cavity using three different Hamiltonians - the coulomb gauge Hamiltonian $\hat{H}_C$, the dipole gauge Hamiltonian, and the quantum optics dipole interaction Hamiltonian $\hat{H}_d$ in Eqs.~\eqref{H_C}, \eqref{H_D-1}, and \eqref{eq:QOdipoleH} as a function of the dimensionless coupling constant $\theta_{\mr{C}} = g_0/\sqrt{\omega_s \omega_c}$. As it is a unitary transformation, the PZW does not affect the eigenvalues of the Hamiltonian, giving us perfect agreement between the spectrum calculated from $\hat{H}_C$ and $\hat{H}_D$. In contrast, the quantum optics dipole interaction Hamiltonian agrees with the minimal coupling Hamiltonian only in the WC regime. As we enter the USC ($\theta_{\mr{C}} > 0.1$) regime, the low energy spectrum, shown in the Fig.~\ref{fig:spectrum_comparison} exhibits disagreement between $\hat{H}_d$ and $\hat{H}_C$. The absence of the diamagnetic term leads to the occurrence of instabilities in $\hat{H}_d$ manifested in the form of complex normal mode frequencies here. These are clearly absent in the full Hamiltonian $\Hop_C$.
	
	\subsection{Numerical Diagonalisation using Bosonic Bogoliubov Transformations}
	
	We present here some additional details of the diagonalization procedure following \cite{key-6}. We want to diagonalize the $2N_{\mr{tot}} \times 2N_{\mr{tot}}$ sized matrix $H$ in Eq.~\eqref{eq:HCMatForm}, using a transformation that preserves the structure of the vector of operators $\bm{\etaop}$ (i.e $\eta_{i+N_{\mr{tot}}}=\eta_{i}^{\dagger}$),
	where the creation operators follow the annihilation operators. For the diagonalization, let us consider a bosonic Bogoliubov transformation \cite{Serafini} given by the matrix $T$ which satisfied 
	\begin{align}
		T^{\dagger}\Lambda T=H,
	\end{align}
	with $\Lambda = \left(D\bigoplus D\right)$ and $D$ representing the diagonal matrix holding the normal mode frequencies i.e. $\lambda_k = D_{kk}$. In order to obtain $T$, we need to first find the unitary matrix $U$, which transforms the matrix $\bar{H}$, defined as
	\begin{align}
		\bar{H}=H^{\frac{1}{2}}\Omega H^{\frac{1}{2}}, \,\,\,\,  \Omega=\left(\begin{array}{cc}
			\mathbb{I} & 0\\
			0 & -\mathbb{I}
		\end{array}\right),
	\end{align}
	into a block diagonal form such that 
	\begin{align}
		U\bar{H}U^{\dagger}=D\bigoplus-D.
	\end{align}
	Using the $U$ and $D$ obtained from this procedure we get
	\begin{align}
		T=\left(D^{-\frac{1}{2}}\bigoplus D^{-\frac{1}{2}}\right)UH^{\frac{1}{2}}.
	\end{align}
	In terms of the transformation $T$, the Hamiltonian can be written as
	\begin{align*}
		H_{C}=\frac{1}{2}\bm{\etaop}^{\dagger}T^{\dagger}\Lambda T\bm{\etaop}-\frac{1}{2} {\rm Tr}\left[H_{d}\right].
	\end{align*}
	Defining new bosonic variables as $\bm{\zetaop}=T \bm{\etaop}$, we can immediately derive the final diagonal form of the Hamiltonian given in Eq.~\eqref{eq:HopDiagonal} of the main text. The evolution of the $\bm{\zetaop}$ operators is simply given by $\hat{\zeta}_{i}\left(t\right)=e^{-i\lambda_{i}t}\hat{\zeta}_{i}\left(0\right)$.
	Thus a two point correlator of the original operators at arbitrary
	time $t$ can be expressed in terms of $\hat{\zeta}$ operators at $t=0$
	via the Bogoliubov transformation $T$. Thus, for the correlator $\left\langle \hat{\eta}_{p}^{\dagger}\hat{\eta}_{q}\right\rangle$,
	we obtain
	\begin{align}
		&\left \langle \etaop_{p}^{\dagger}\etaop_{q}\right\rangle (t) = \sum_{l=1}^{N_{\mr{tot}}}\sum_{m=1}^{N_{\mr{tot}}}\left\{ \Gamma_{qm}^{lp}\left\langle \zetaop_{l}^{\dagger}(0)\zetaop_{m}(0)\right\rangle e^{i\left(\lambda_{l}- \lambda_{m}\right)t} \right . \nonumber\\
		& \left .  +\Gamma_{q(m+N_{\mr{tot}})}^{lp}\left\langle \zetaop_{l}^{\dagger}(0)\zetaop_{m}^{\dagger}(0)\right\rangle e^{i\left(\lambda_{l}+\lambda_{m}\right)t}\right\} \nonumber \\
		& +\sum_{l=1}^{N_{\mr{tot}}}\sum_{m=1}^{N_{\mr{tot}}}\left\{ \Gamma_{qm}^{(l+N_{\mr{tot}})p}\left\langle \zetaop_{l}(0)\zetaop_{m}(0)\right\rangle e^{-i\left(\lambda_{l}+\lambda_{m}\right)t} \right . \nonumber \\
		& \left . + \Gamma_{q(m+N_{\mr{tot}})}^{(l+N_{\mr{tot}})p}\left\langle \zetaop_{l}(0)\zetaop_{m}^{\dagger}(0)\right\rangle e^{-i\left(\lambda_{l}-\lambda_{m}\right)t}\right\}     \label{eq:twopointcorrdyn}
	\end{align}
	where $\Gamma_{qm}^{lp}=A_{lp}^{\dagger}A_{qm}$, $A=T^{-1}$ Using the transformation $\bm{\zetaop}=T\bm{\etaop}$, the two point correlators of $\bm{\zetaop}$ at $t=0$ can be determined from the two-point correlations of the original $\bm{\etaop}$ operators at $t=0$ after the initial state of the system is specified.
	
	\section{Additional Details on the Weak Coupling Regime}
	\label{app:B}
	\setcounter{figure}{0}
	\renewcommand{\thefigure}{B\arabic{figure}}
	
	In this appendix we provide some additional details regarding the weak coupling regime results discussed in Sec.~\ref{sec:WeakIntMarkov}. In particular, we detail how we handle some subtle issues that arise in the choice of the EM environment parameters as well as additional coarse graining that we need to perform in order to match the exact dynamics for a finite sized reservoir with the predictions of the Lindblad master equation.
	
	\subsection{Choice of Reservoir parameters for numerical convergence}
	
	\subsubsection{Choice of ultraviolet cutoff for the cavity EM environment}
	
	For the ideal 1-D cavity EM environment we consider with the spectrum $\omega_n^R = n c \pi/L$, there is no upper-bound on the allowed frequency of the reservoir modes. As discussed in the main text, the exact numerics have to performed by assuming a finite number of environment modes and hence introducing an ultraviolet cut-off $\omega_{\mr{uv}}$. While the qualitative features of the dynamics remain the same once $\omega_{\mr{uv}}$ is chosen to be much larger than the system frequency $\omega_s$, the exact quantitative dynamics changes as we change $\omega_{\mr{uv}}$ but converges eventually to a cut-off independent behavior. In particular in the weak coupling regime calculations in Sec.~\ref{sec:WeakIntMarkov} our choice of $\omega_{\mr{uv}}=2\,\omega_s$ leads to well-converged dynamics. Note also that this choice places the frequency of the system dipole oscillator at the center of the reservoir spectrum. We are able to obtain convergence with a fairly small value of $\omega_{\mr{uv}}$ in this case since as illustrated in Fig.~\ref{fig:polariton_pop}(a), only the modes near resonance contribute to the dynamics. On the other hand, as evident from Fig.~\ref{fig:polariton_pop}, increasing the interaction strength leads to a larger spread around the resonant normal modes in the USC regime. Consequently, in the calculations in the USC regime in Sec.~\ref{sec:stronginteraction} we require a larger value of $\omega_{\mr{uv}}=10\,\omega_s$ for convergence. Finally, in the AdC regime since the dynamics is again dominated by the lowest polariton mode, the dynamics converge for smaller values of $\omega_{\mr{uv}} = 2 \omega_s$.
	
	\subsubsection{Choice of hopping parameter in a coupled cavity array}
	
	Unlike the ideal cavity which has no upper bound on the allowed reservoir mode frequency, the cavity array has a bounded spectrum. The number of modes used in a numerical calculation is then set by the number of cavities or length of the array. Our task is then to choose the length of the cavity array $N$ and the hopping parameter $J$ such that the population dynamics at weak coupling gives us the exponential decay predicted by the Lindblad equation. Throughout our considerations in Sec.~\ref{sec:WeakIntMarkov} for the cavity array EM environment, we keep the dipole oscillator frequency $\omega_s$ fixed and equal to the array on-site energy $\omega_c$. In this case, choosing $J/\omega_c$ large leads to a large propagation velocity $v_s = Ja$ in the array and consequently makes it hard to satisfy condition Eq.~\eqref{eq:MarkovcondCAform1}. On the other hand, taking $J/\omega_c \ll 1$ leads to a very flat spectrum and a diverging density of states $D(\omega=\omega_c) \propto 1/J$ at the dipole frequency. In a sense, in the $J \rightarrow 0$, the density of states at the band center and band edge are both divergent and consequently the non-Markovian behavior pointed out in \cite{PhysRevA.96.043811} becomes pronounced even within the band. Thus, avoiding these two extremes, in our exact calculations we choose $J/\omega_c = 0.5$ which gives excellent agreement with the Lindblad dynamics in the weak coupling regimes.
	
	\subsection{Coarse Graining Protocol}
	
	The Markovian Lindblad master equation describes the dynamics of a system interacting with a reservoir over a coarse-grained time scale \cite{Cohen} that is much larger than the short time-scales over which system-reservoir and intra-reservoir correlations build up. On the other hand, the exact numerical simulation we perform here can track the dynamics of the system and reservoir over the latter short time-scales. Thus, to provide a faithful comparison between the exact numerical solution and the Lindblad master equation, apart from ensuring that we are in the weak coupling regime we need to do an additional coarse graining of the numerical results. To this end in the WC regime presented in Sec.~\ref{sec:WeakIntMarkov}, we perform a rolling average of the results from the exact numerical solution over a time interval equal $t_{\mr{exc}}$ in the results depicted in Figs.~\ref{fig:1e_rad}-\ref{fig:10e_rad_lam10} in Sec.~\ref{sec:WeakIntMarkov}. A reliable estimate of $t_{\mr{exc}}$ is obtained by identifying the time scale of the fast oscillations in the cavity and cavity array dynamics. For the cavity reservoir, we find that the Rabi frequency of exchange between the dipole and the fundamental cavity mode, given by
	\begin{align*}
		\Omega_{\mr{C}}= & \omega_s\sqrt{1+\theta_{\mr{C}}^{2}}
	\end{align*}
	provides a good estimate. In the WC regime, the $\theta_\mr{C}$-dependence is negligible and the timescale of this Rabi exchange is approximated as $t_{\mr{exc}} \simeq  \frac{2\pi}{\omega_s}$. For the cavity array case, the fastest frequency corresponds to the exchange between the dipole and the reservoir mode corresponding to $k=0$, given by
	\begin{align*}
		\Omega_{\mr{CA}}=  J\sqrt{1+\kappa\theta_{\mr{CA}}^{2}},
	\end{align*}
	with $\kappa =\frac{\pi a}{NJ\left(1-\frac{J}{\omega_{c}}\right)}$.
	For our choice of $J/\omega_c = 0.5$, the $\theta_{\mr{CA}}^2$-dependence can again be neglected at weak coupling and the timescale of exchange is approximated by $t_{\mr{exc}}\simeq  \frac{2\pi}{J}$. As we see in the results presented in Sec.~\ref{sec:WeakIntMarkov}, this coarse graining process removes the fast oscillations from the dynamics and gives us good agreement with the predictions of the Lindblad Master equation.
	
	\subsection{Additional figures for dipole population dynamics}
	
\begin{figure}
		\begin{centering}
			\subfloat{\begin{overpic}[abs,width=0.8\columnwidth]{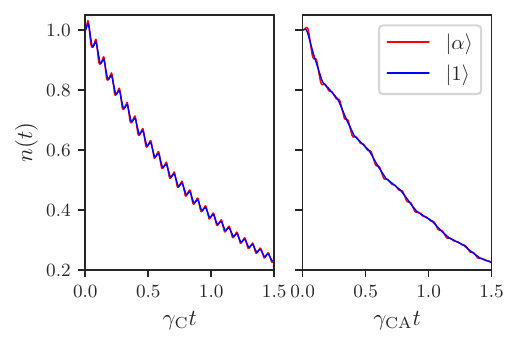}
					\put(40,50){\textbf{(a)}}
					\put(118,50){\textbf{(b)}}
				\end{overpic}
			}\\
		\end{centering}
		\caption{Spontaneous decay of single dipole oscillator coupled weakly to an ideal cavity (a) and a cavity array reservoir (b). The red and blue lines represent the population dynamics with the system dipole initially in a coherent and Fock state respectively with average initial photon number of $1$. System and reservoir parameters are the same as Fig.~\ref{fig:1e_rad}(a) and Fig.~\ref{fig:CCA_dec_1e}(a) respectively.}
		\label{fig:1e_coherent}
	\end{figure}

	Let us finally present some additional results in the weak coupling regime complementing those shown in Sec.~\ref{sec:WeakIntMarkov}. Fig. \ref{fig:1e_coherent} shows the single dipole population dynamics for Fock and coherent initial states in a cavity and cavity array at $\phi_{\mr{C/CA}}=0.05$. It can be observed that the population dynamics in the Markovian regime corresponds to similar exponential decays for both the Fock and the coherent initial state in the cavity as well as cavity array environment. 
	
\begin{figure}
		\begin{centering}
			\subfloat{\begin{overpic}[abs,width=0.8\columnwidth]{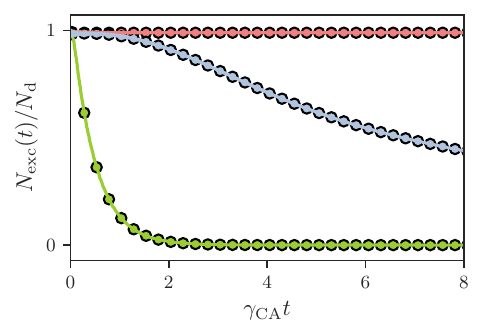}
					\put(128,55){\textbf{(a)}}
			\end{overpic}}\\
			\subfloat{\begin{overpic}[abs,width=0.8\columnwidth]{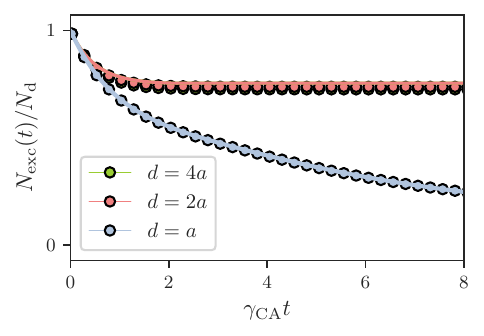}
					\put(128,55){\textbf{(b)}}
			\end{overpic}}
		\end{centering}
		\caption{Total excitation dynamics of four oscillators placed symmetrically about the center at the sites of a cavity array reservoir with $\phi_{\mr{CA}}=0.02$ in coherent (a) and Fock (b) initial states for different separations $d$. Results from the coarse grained exact numerical calculations are (circles) compared with the Lindblad master equation results (solid lines). System and reservoir parameters are kept the same as Fig.~\ref{fig:CCA_dec_1e}(a). \label{fig:4e_CCA_pop}}
	\end{figure}

\begin{figure}
		\begin{centering}
			\subfloat{\begin{overpic}[abs,width=0.8\columnwidth]{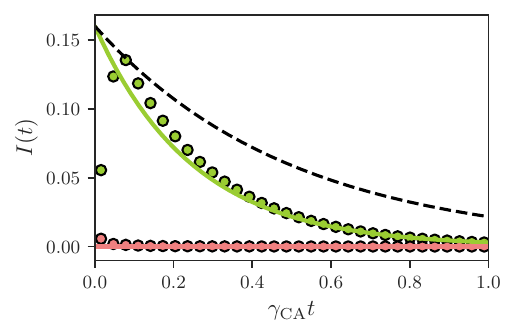}
					\put(95,75){\textbf{(a)}}
			\end{overpic}}\\
			\subfloat{\begin{overpic}[abs,width=0.8\columnwidth]{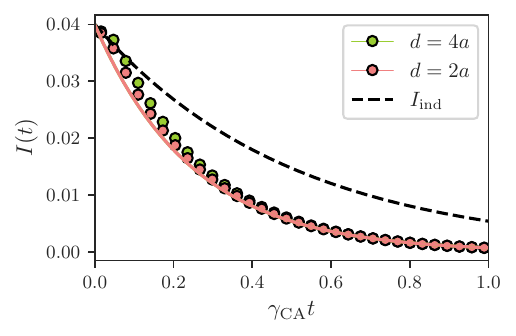}
					\put(95,75){\textbf{(b)}}
			\end{overpic}}
		\end{centering}
		\caption{Radiation rate of four oscillators placed symmetrically about the center at the sites of a cavity array reservoir with $\phi_{\mr{CA}}=0.02$ in coherent (a) and Fock (b) initial states for different separations $d$. Results from the coarse grained exact numerical calculations (circles) are  compared with the Lindblad master equation results (solid lines) and scaled independent oscillator dynamics (dashed lines). Other parameters are kept the same as in Fig.~\ref{fig:CCA_dec_1e}(a).\label{fig:4e_CCA_rad}}
	\end{figure}

\begin{figure}
		\begin{overpic}[abs,width=0.8\columnwidth]{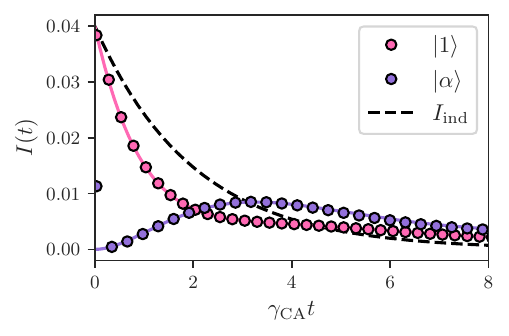}
		\end{overpic}
		\caption{Radiation rate of four oscillators placed symmetrically about the center at the sites of a cavity array reservoir inside a cavity with separation $d=a$ and
			$\phi_{\mr{CA}}=0.02$ for coherent and Fock initial states. Results from the coarse grained exact numerical calculations (circles) are compared with the Lindblad master equation results (solid lines) and scaled independent oscillator dynamics (dashed lines). Other parameters are kept the same as Fig.~\ref{fig:CCA_dec_1e}(a).\label{fig:4e_rad_lam10}}
	\end{figure}

	Coming to arrays of dipole oscillators, we presented the collective dynamics results for the cavity EM environment in the main paper. Here the collective population dynamics of four dipoles placed at various separations in a cavity array reservoir is shown in Fig.~\ref{fig:4e_CCA_pop} for (a) coherent and (b) Fock initial states. The behavior is qualitatively similar to the collective decay inside an ideal cavity shown in Fig. \ref{fig:10e_pop}. This similarity is also observed in the plots for the radiation rate dynamics inside the cavity array plotted in Figs.~\ref{fig:4e_CCA_rad} and \ref{fig:4e_rad_lam10} when comparing to the cavity reservoir case shown in Fig.~(\ref{fig:10e_rad}) and Fig.~(\ref{fig:10e_rad_lam10}).
	
	\section{Additional Details on the Strong Coupling Regimes}
	\label{app:C}
	\setcounter{figure}{0}
	\renewcommand{\thefigure}{C\arabic{figure}}
	\subsection{Calculation of \texorpdfstring{$g_\mr{eff}$}{geff} for cavity array}
	Unlike the ideal cavity where the reservoir frequencies have a linear dependence on $k_n$, the frequencies of the cavity array reservoir are non-linearly spaced. Hence the $n$-dependence from $\omega^R_n  =\omega_c - J\cos(k_n a)$  cannot be pulled out of the expression for $\bar{g}_n$ in the simple manner that is used for the ideal cavity reservoir. For the purpose of parameterizing the interaction strength, we require a quantity that is independent of the reservoir size $N$. To this end, we define an effective coupling strength as the sum of the bare couplings to the reservoir modes in $k$-space  
	\begin{align}
		g_{\text{eff}}^{2}= & \sum_{n}\bar{g}_{n}^{2}=\sum_{n=-\frac{N}{2}}^{\frac{N}{2}}\frac{g_{0}^{2}\omega_s}{\omega_s-J\cos\left(k_{n}a\right)}
	\end{align}
	Here we have considered only the zero-detuning case, i.e. $\omega_c = \omega_s$. $k_n$ can be written in terms of $n$ as $k_n=2\pi n/(Na)$  where  $n\in\mathbb{Z}$ and $-N/2<n\leq N/2$. In the large $N$ limit, this sum maybe approximated as an integral over $k$. Under a substitution of $k'=ka$ as the variable of integration, the effective coupling strength can be expressed as
	\begin{align}
		g_{\text{eff}}^{2} \simeq & \frac{\bar{g}_{0}^{2}}{\pi}\left(\frac{\omega_s}{Ja}\right)\intop_{0}^{+\pi}\frac{dk'}{\left[\frac{\omega_s}{J}-\cos\left(k'\right)\right]}.
	\end{align}
	The above integral is solvable when $\omega_s/J > 1$ as the solution is given by $\int_{0}^{\pi}\left(p-\cos x\right)^{-1}dx=\pi/\sqrt{p^{2}-1}$. Thus the integral introduces a factor of $\pi/\sqrt{\left( \omega_s/J\right)^{2}-1}$ to $g_{\text{eff}}^2$. This factor becomes simply a multiplicative constant of order $\sim 2$ for the results presented in Sec.~\ref{sec:stronginteraction} where $J/\omega_s=0.5$ is chosen. 
	\subsection{Uncoupled modes in the polariton spectrum for a single dipole}
	In Figs.~\ref{fig:classification} and \ref{fig:spectrum_CCA} we have plotted the normal mode or polariton spectrum for a single dipole interacting with an ideal cavity and cavity array reservoir, respectively. There, we pointed out that for the dipole placed at the center of the cavity or array, there are always a collection of modes that are uncoupled. We want to identify these modes here. 
	
	From the mode function for the ideal cavity in Eq.~\eqref{eq:ModeFnCavity}, it is immediately clear for the oscillator placed at the center of the cavity, the amplitude of all the even $n$ standing waves (with $f_n(x) \propto \sin (k_n x)$) vanishes at $x=0$. Hence the frequency of these uncoupled reservoir modes, which represent half of the modes in our numerical calculation given by $\omega_n = n \omega_c$ ($n$ even), remains unaffected by the dipole and is shown by dashed white lines in Fig.~\ref{fig:classification}. 
	
	For the cavity array reservoir the mode functions are given by Eq.~\eqref{eq:ModeFnCavityArray} and are proportional to $e^{\pm i k_n x}$. Thus it does seem like all the modes are coupled to the dipole oscillator placed at $x=0$. But, as we show now, by exploiting the degeneracy of the modes with $k_n = \pm n \frac{2\pi}{L}$ we find that half of the modes are still decoupled from the dipole oscillator. In order to see this let us first  decompose the exponential mode function in Eq.~\eqref{eq:ModeFnCavityArray} into sines and cosines and write down the corresponding transformed reservoir operators as
	\begin{align}
		\Rop_{1n}= & \frac{\Rop_{n}+\Rop_{-n}}{\sqrt{2}} & \Rop_{2n}= & \frac{\Rop_{n}-\Rop_{-n}}{\sqrt{2}}.
	\end{align}
	With this transformation gives the system-reservoir interaction in the Hamiltonian Eq.~\eqref{H_C} reduces to the form
	\begin{align}
		-i\sum_{n=\frac{1}{2}}^{+\frac{N}{2}}g_{0}\sqrt{\frac{2\omega_{s}}{\omega_{n}^{R}}}\left(\aop^{\dagger}-\aop\right)\cdot & \left\{ \cos\left(k_{n}x\right)\left(\Rop_{1n}+\Rop_{1n}^{\dagger}\right)\right. \label{eq:CAmodefncoup} \\
		& \left.+i\sin\left(k_{n}x\right)\left(\Rop_{2n}^{\dagger}-\Rop_{2n}\right)\right\} \nonumber. 
	\end{align}
	With the above form it is immediately clear that half of the reservoir modes are coupled via a $\sin(k_n x)$ coupling, which vanishes when the dipole is placed at the site with $x=0$. These uncoupled modes are again distinguished with dashed white lines in Fig.~\ref{fig:spectrum_CCA}. Moreover, in the weak coupling regime the spatial distribution of photon population in polariton Fock states displayed in Fig.~\ref{fig:Polariton_field_CCA} (a,b) for $k_n \neq 0$ are periodic as they reflect the mode function corresponding to $\hat{R}_{1n}$ in Eq.~\eqref{eq:CAmodefncoup}. 
	\subsection{Reservoir dynamics and examples of array of dipoles strongly coupled to environment}
	\begin{figure}[t]
		\begin{centering}
			\begin{overpic}[abs,width=0.9\columnwidth]{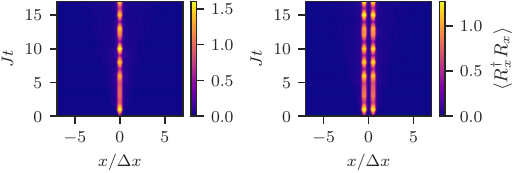}
				\put(30,30){\textbf{\textcolor{white}{(a)}}}
				\put(138,30){\textbf{\textcolor{white}{(b)}}}
			\end{overpic}
		\end{centering}
		\caption{Photon population dynamics at the sites of a coupled cavity array for a single dipole in an initial Fock state (a) and two dipoles in the initial state $\ket{\psi_{-}}$ with a separation of $\Delta x = 2a$ (b) strongly coupled to a cavity array ($\theta_{\mr{CA}} = 10^{-0.1} $). Other parameters are kept the same as in Fig.~\ref{fig:1e_USCweak} \label{fig:Bath_dyn_CCA}}
	\end{figure}
	\begin{figure}
		\begin{centering}
			\begin{overpic}[abs,width=0.8\columnwidth]{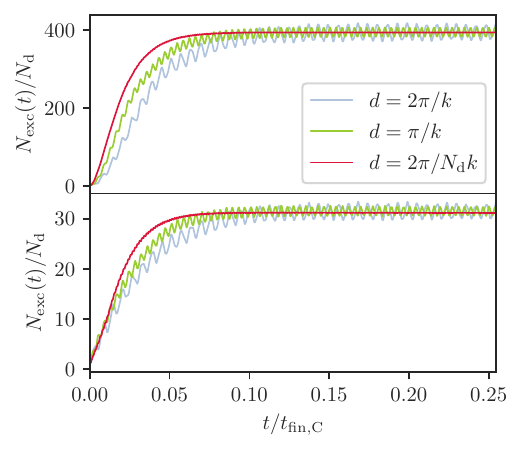}
				\put(60,40){\textbf{(b)}}
				\put(60,110){\textbf{(a)}}
			\end{overpic}
		\end{centering}
		\caption{Total excitation dynamics of $10$ oscillators placed symmetrically about the center of an ideal cavity cavity in the USC regime with $\theta_{\mr{C}}=10^{-0.1}$ in coherent (a) and Fock (b) initial states for different separations $d$. Other parameters are kept the same as in Fig.~\ref{fig:1e_USCweak}.\label{fig:10e_pop_USC}}
	\end{figure}
	The large number of polaritons that are occupied in the USC regime as well as the requirement of a large ultraviolet cutoff makes the calculation of the reservoir dynamics computationally expensive. However the fixed number of sites in the cavity array makes the calculation of USC reservoir dynamics computationally easier as compared to the ideal cavity case. Fig.~\ref{fig:Bath_dyn_CCA} shows the dynamics of reservoir site populations $\langle \hat{R}_x^\dagger \hat{R}_x \rangle$ for a single dipole initially in a Fock state [(a)] and two dipoles initially in a subradiant state [(b)]. It is observed that in contrast to the WC reservoir dynamics (see Fig.~\ref{fig:reservoir_dyn}), there is no spatial propagation of light through the sites of the cavity array reservoir. The light emitted by the dipole is seen pulsing at the cavity sites where the dipoles are located within the reservoir, indicating the exchange which gives us oscillatory population dynamics for the dipole. As evident from Fig.~\ref{fig:Bath_dyn_CCA} (b), this behavior is also seen in the case of two dipoles inside the cavity array reservoir.
	
	Fig.~\ref{fig:10e_pop_USC} shows the dynamics of the total excitation number $N_{\mr{exc}}(t)$ for 10 dipole oscillators ultra-strongly coupled to a cavity in the USC regime. The dipoles are placed symmetrically about the center of the cavity at separations of $\lambda$, $\lambda/2$ and $\lambda/10$. Unlike the qualitatively different dynamics we saw at different separations for the collective dissipation of 10 dipoles weakly coupled to a cavity (see Fig.~\ref{fig:10e_pop}), the USC collective dynamics do not exhibit any qualitative separation dependence. This absence of separation dependence is observed in the USC collective dynamics for both the coherent and Fock initial states shown in Figs.~\ref{fig:10e_pop_USC}(a) and (b) respectively.
	%
	
\end{document}